\documentclass[twocolumn,secnumarabic,amssymb, nobibnotes, aps, prl,superscriptaddress]{revtex4-1}

\usepackage{amsmath}
\usepackage{amsfonts}
\usepackage{amssymb}
\usepackage{graphicx}
\usepackage{hyperref}
\usepackage{longtable}
\usepackage{dcolumn}
\usepackage{bm}
\usepackage{epstopdf}
\usepackage{color}

\begin{document}
\title{Shear-driven instabilities of membrane tubes and dynamin-induced scission}

\author{Sami C.~Al-Izzi}
\thanks{Current affiliation: School of Physics \& EMBL-Australia node in Single Molecule Science, University of New South Wales, Sydney, Australia}
\affiliation{Department of Mathematics, University of Warwick, Coventry CV4 7AL, UK}
\affiliation{Department of Physics, University of Warwick, Coventry CV4 7AL, UK}
\affiliation{Institut Curie, PSL Research University, CNRS, Physical Chemistry Curie, F-75005, Paris, France}
\affiliation{Sorbonne Universit\'{e}, CNRS, UMR 168, F-75005, Paris, France}
\author{Pierre Sens}%
\affiliation{Institut Curie, PSL Research University, CNRS, Physical Chemistry Curie, F-75005, Paris, France}
\affiliation{Sorbonne Universit\'{e}, CNRS, UMR 168, F-75005, Paris, France}
\author{Matthew S.~Turner}%
\affiliation{Department of Physics, University of Warwick, Coventry CV4 7AL, UK}
\affiliation{Centre for Complexity Science, University of Warwick, Coventry CV4 7AL, UK}

\begin{abstract}
Motivated by the mechanics of dynamin-mediated membrane tube fission we analyse the stability of fluid membrane tubes subjected to shear flow in azimuthal direction. We find a novel helical instability driven by the membrane shear flow which results in a non-equilibrium steady state for the tube fluctuations. This instability has its onset at shear rates that may be physiologically accessible under the action of dynamin and could also be probed using \textit{in-vitro} experiments on membrane nanotubes, e.g. using magnetic tweezers. We discuss how such an instability may play a role in the mechanism for dynamin-mediated membrane tube fission.
\end{abstract}

\maketitle

The covariant hydrodynamics of fluid membranes has been a subject of much interest in the soft matter and biological physics community in recent years, both for the general theoretical features of such systems \cite{cai_covariant_1994,cai_hydrodynamics_1995,fournier_hydrodynamics_2015,sahu_irreversible_2017,sahu_geometry_2020} and their application to biological processes \cite{sens_dynamics_2004,arroyo_relaxation_2009,brochard-wyart_hydrodynamic_2006,morris_mobility_2015}. Such systems couple membrane hydrodynamics with bending elasticity and have been shown to  display complex visco-elastic behaviour in geometries with high curvature \cite{rahimi_curved_2013}.

Membrane tubes are highly curved and are found in many contexts in cell biology, including the endoplasmic reticulum and the necks of budding vesicles \cite{kaksonen_mechanisms_2018}. Such tubes can be pulled from a membrane under the action of a localized force (such as from molecular motors) \cite{derenyi_formation_2002,yamada_catch-bond_2014,cuvelier_coalescence_2005}. They are stable due to a balance between bending energy, involving the bending rigidity $\kappa$, and  the surface tension $\sigma$ with an equilibrium radius $r_0=\sqrt{\frac{\kappa}{2\sigma}}$ \cite{zhong-can_bending_1989}. 

\begin{figure}[h!]
\includegraphics[width=0.48\textwidth,trim = 0cm 4cm 0cm 0cm,clip=true]{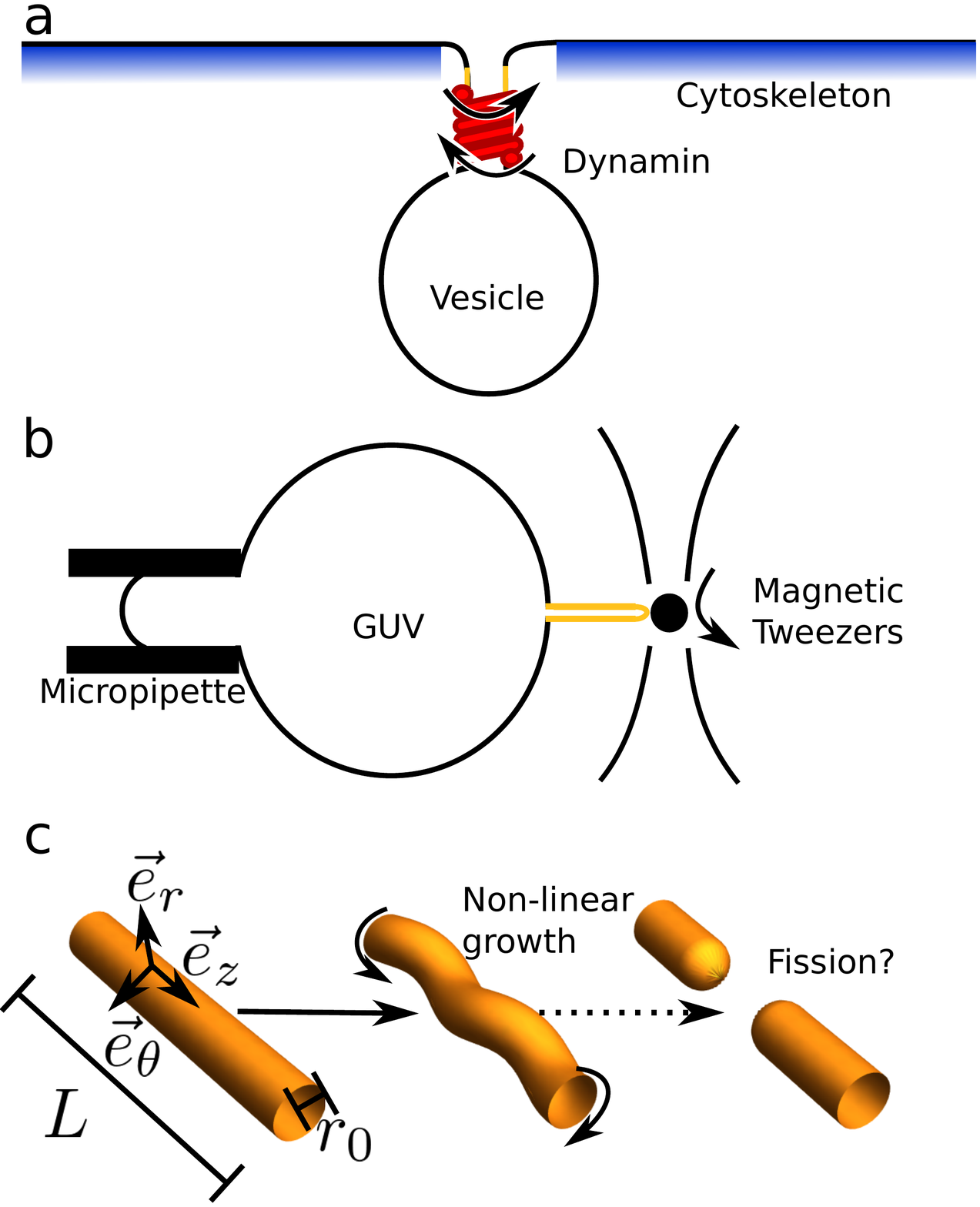}
\caption{\label{fig:diagram}Possible realizations of shear driven instabilities on membrane tubes (shown in orange throughout). a) Dynamin on the neck of a budding vesicle. The protein constricts and (counter)rotates, prior to fission of the tube. This rotation drives a significant shear flow near the neck of the vesicle. We discuss the possible effects of the Gaussian curvature of the neck in the conclusion. b) A GUV with membrane tube pulled by magnetic tweezers; the magnetic bead can be spun in order to drive flows in the azimuthal direction on the tube. c) Sketch of the growth of the helical instability described in this letter, the final stage is a possible pathway to tube fission due to non-linear effects. The basis vectors on the membrane $\vec{e}_i$ where $i=r,\theta,z$, length of tube, $L$, and equilibrium radius, $r_0$, are labelled. Middle panel shows shear direction.}
\end{figure}

One of the simplest ways to drive flows on the surface of these tubes is to impose a velocity in the azimuthal direction. The analysis of shape changes induced by such flows is the subject of this letter. Two possible mechanisms for realizing such flows via \textit{in-vitro} and \textit{in-vivo} experiments are shown in Fig.~\ref{fig:diagram}.

The fission of membrane tubes plays an important role in many cellular processes, ranging from endocytosis to mitochondria fission \cite{mcclure_dynamin_1996,frank_role_2001}. The key component of the biological machinery required to induce membrane fission is a family of proteins called dynamin that hydrolyse GTP into GDP \cite{antonny_membrane_2016,roux_gtp-dependent_2006}. Dynamin is a protein complex that oligomerizes to form polymers which wrap helically around membrane tubes \cite{antonny_membrane_2016,roux_membrane_2010,shlomovitz_membrane-mediated_2011}. Although there is clear evidence that dynamin undergoes a conformational change when it hydrolyses GTP, there is not yet a consensus on the exact method of fission \cite{roux_reaching_2014,kozlov_dynamin_1999,kozlov_fission_2001,mcdargh_constriction_2016,mcdargh_dynamins_2018}.  Recent coarse-grained simulations have shed some light on the possible role of constriction and de-polymerisation \cite{pannuzzo_role_2018}.
 It has been shown experimentally that, upon hydrolysis of GTP, dynamin (counter)rotates rapidly whilst constricting \cite{roux_gtp-dependent_2006}, giving a mechanism for the generation of flows in the azimuthal direction. Another possible way of driving such flows is by pulling a narrow membrane (nano)tube from either a Giant Unilamellar Vesicle (GUV) or a cell using magnetic tweezers. Magnetic field oscillations can then be used to spin the attached magnetic bead \cite{crick_physical_1950,hosu_eukaryotic_2007,monticelli_magnetic_2016}, thereby setting up a frictional flow in the tube.

The membrane behaves as a viscous fluid with $2$D viscosity $\eta_m$. The Saffman-Delbr\"uck length, $L_{SD}=\frac{\eta_m}{\eta}$ \cite{saffman_brownian_1975,saffman_brownian_1976,henle_hydrodynamics_2010}, with $\eta$ the bulk  fluid viscosity, is the distance over which bulk hydrodynamics screens membrane flows in planar geometry. In the case of a membrane tube the screening length is modified due to geometric effects and becomes $\sqrt{L_{\text{SD}}r_0}$, \cite{henle_hydrodynamics_2010}. We  consider dynamics on a scale less than this, such that the dominant dissipation mechanism involves the membrane flows. This means that we can neglect bulk flows on sufficiently short length-scales (short tubes) \cite{morris_mobility_2015,bahmani_analysis_2016}. For further details see S.I.

We consider a lipid membrane as a manifold equipped with metric $g_{ij}$ and second fundamental form $b_{ij}$ \cite{frankel_2011}. The coordinate basis is defined by the triad $\{\vec{e}_1,\vec{e}_2,\vec{n}\}$ where $\vec{e}_i$ and $\vec{n}$ are the basis of the tangent bundle and normal bundle of the surface respectively. The surface has velocity, $\vec{V}=\boldsymbol{v} + w\vec{n}$ where $\boldsymbol{v}=v^i\vec{e}_i$. We label vectors in the membrane tangent space in bold, e.g.~$\boldsymbol{x}$, and vectors in $\mathbb{R}^3$ with arrows, e.g.~$\vec{x}$. We define the mean and Gaussian curvature as $2H=b^i{}_i$ and $K=\det{b_i{}^{j}}$ respectively. We assume the membrane behaves like a zero-Reynolds number fluid in the tangential direction \cite{happel_low_1983} and has bending energy given by the usual Helfrich energy \cite{helfrich_elastic_1973}. Surface tension, $\sigma$, is treated as a Lagrange multiplier imposing membrane area conservation. We will assume zero spontaneous curvature for simplicity. For conciseness we will simply state the equations of motion for the membrane, for details on their derivation see \cite{arroyo_relaxation_2009,rangamani_interaction_2013} or S.I. 

The continuity equation for an incompressible membrane is given by
\begin{equation}\label{eq:continuity}
\nabla_iv^i=2H w
\end{equation}
which is simply the Euclidean continuity equation modified to account for the normal motion of the membrane \cite{arroyo_relaxation_2009,marsden_mathematical_1994}.

Force balance normal to the membrane means the normal elastic and viscous forces must sum to zero, leading to the following
\begin{equation}\label{eq:shape}
\begin{split}
\kappa\left[2\Delta_{\text{LB}}H -4H\left(H^2-K\right)\right] +2\sigma H\\ + 2\eta_m \left[b^{i}{}_j\nabla_i v^j- 2\left(2H^2-K\right)w\right]=0
\end{split}
\end{equation}
Here $\kappa$ is the bending rigidity of the membrane and $\Delta_{\text{LB}}$ is the Laplace-Beltrami operator. Note that we are using a geometrical definition of $\Delta_{\text{LB}}$ that is analogous to a curl-curl operator on a manifold, hence the sign difference with the usual Laplacian operator in the shape equation (see S.I. for details). This is a modified form of the shape equation first derived by Zhong-Can \& Helfrich \cite{zhong-can_bending_1989}, but with the addition of viscous normal forces given by fluid flow on the membrane. The term coupling the second fundamental form and gradients in tangential velocity can be thought of as the normal force induced by fluid flowing over an intrinsically curved manifold. This term is of fundamental importance in the present study as it drives a shape instability. The other non-standard term  $\sim\left(2H^2-K\right)w$ is the dissipative force associated with the normal velocity, inducing flows in the tangential direction on a curved surface.

Force balance in the tangential direction gives
\begin{equation}\label{eq:2dStokesFull}
\begin{split}
\eta_m\left[ \Delta_{\text{LB}}v^i -2Kv^i +2\left(b^{ij}-2H g^{ij}\right)\nabla_j w\right]\\ - \nabla^i\sigma =0
\end{split}
\end{equation}
which is the modified form of the $2$D Stokes equations. The new terms, coupling Gaussian curvature with tangential velocity, and curvature components with the gradients in normal velocity, come from the modified form of the rate-of-deformation tensor which accounts for the curved and changing geometry of the membrane. The term $\sim Kv^i$ describes the convergence/divergence of streamlines on a curved surface. The term $\sim\left(b^{ij}-2H g^{ij}\right)\nabla_j w$ describes the forces induced tangentially by the dynamics of the membrane.

We consider a ground-state membrane tube ($w=0$) of length $L$  in cylindrical coordinates $(r,\theta,z)$ with radius $r_0=\sqrt{\frac{\kappa}{2\sigma_0}}$ and impose a velocity $v=v_0\vec{e}_\theta$ at $z=0$ (which can be interpreted as the edge of an active dynamin ring, for example). Making use of the azimuthal symmetry the continuity and Stokes equations reduce to an ODE that admits the solution
\begin{equation}\label{eq:groundState}
\boldsymbol{v}^{(0)}=\left(v_0 - \Omega z\right)\frac{1}{r_0}\vec{e}_\theta
\end{equation}
where the exact value of the shear flow $\Omega$ depends on the boundary condition at $z=L$, but roughly scales as $\Omega\sim\frac{v_0}{L}$ if we either implement torque balance, e.g. at the boundary where a tube joins onto a planar membrane, or simply set $v(L)=0$, see S.I. for more details.

We can now make a perturbation about this ground state in $r(z,\theta,t)=r_0+u(\theta,z,t)$, $\boldsymbol{v}=\boldsymbol{v}^{(0)}+\delta v^{\theta}(\theta,z,t)\vec{e}_\theta+\delta v^{z}(\theta,z,t)\vec{e}_z$, $\sigma=\sigma_0+\delta\sigma(\theta,z,t)$ and $w=\partial_t u$. Making use of the discrete Fourier transform, $f(\theta,z,t)=\sum_{q,m}\bar{f}_{q,m}(t)e^{\dot{\imath}qz+\dot{\imath}m\theta}$, where $\bar{f}_{q,m}$ is the Discrete Fourier Transform of $f(\theta,z)$ with $m\in\mathbb{Z}$ and $q=\frac{2\pi n}{L}$ where $n\in\mathbb{Z}\setminus\{0\}$, we can write Eqs.~\ref{eq:continuity}, \ref{eq:shape}, \ref{eq:2dStokesFull} in Fourier space and linearise in the perturbations. The linear response of the normal force balance is the following
\begin{equation}\label{eq:linearResponseShapeTerm}
\begin{split}
\mathcal{F}^{u}_{q,m}\bar{u}_{q,m} +\mathcal{F}^{\sigma}_{q,m}\bar{\delta\sigma}_{q,m} + \mathcal{F}^{\theta}_{q,m}\bar{\delta v^{\theta}}_{q,m}\\ +\mathcal{F}^{z}_{q,m}\bar{\delta v^{z}}_{q,m} +\mathcal{G}_{q,m}\bar{\delta w}_{q,m}=0
\end{split}
\end{equation}
where $\mathcal{F}^u_{q,m} = \frac{4\sigma_0^2}{\kappa}\big[\tilde{q}^4 +m^4 +2\tilde{q}^2m^2 -2 m^2 +1\big] - \frac{2\eta_m m \tilde{q} \Omega}{r_0^2}$, 
$\mathcal{F}^{\sigma}_{q,m} = \frac{1}{r_0}$, $\mathcal{F}^\theta_{q,m} = \frac{2\dot{\imath} m\eta_m}{r_0^2}$, $\mathcal{F}^z_{q,m} =0$ and $\mathcal{G}_{q,m} = \frac{2\eta_m}{r_0^2}$ where $\tilde{q}=qr_0$.

\begin{figure}
\includegraphics[width=0.4\textwidth,trim = 0.5cm 2cm 0.5cm 2cm, clip=true]{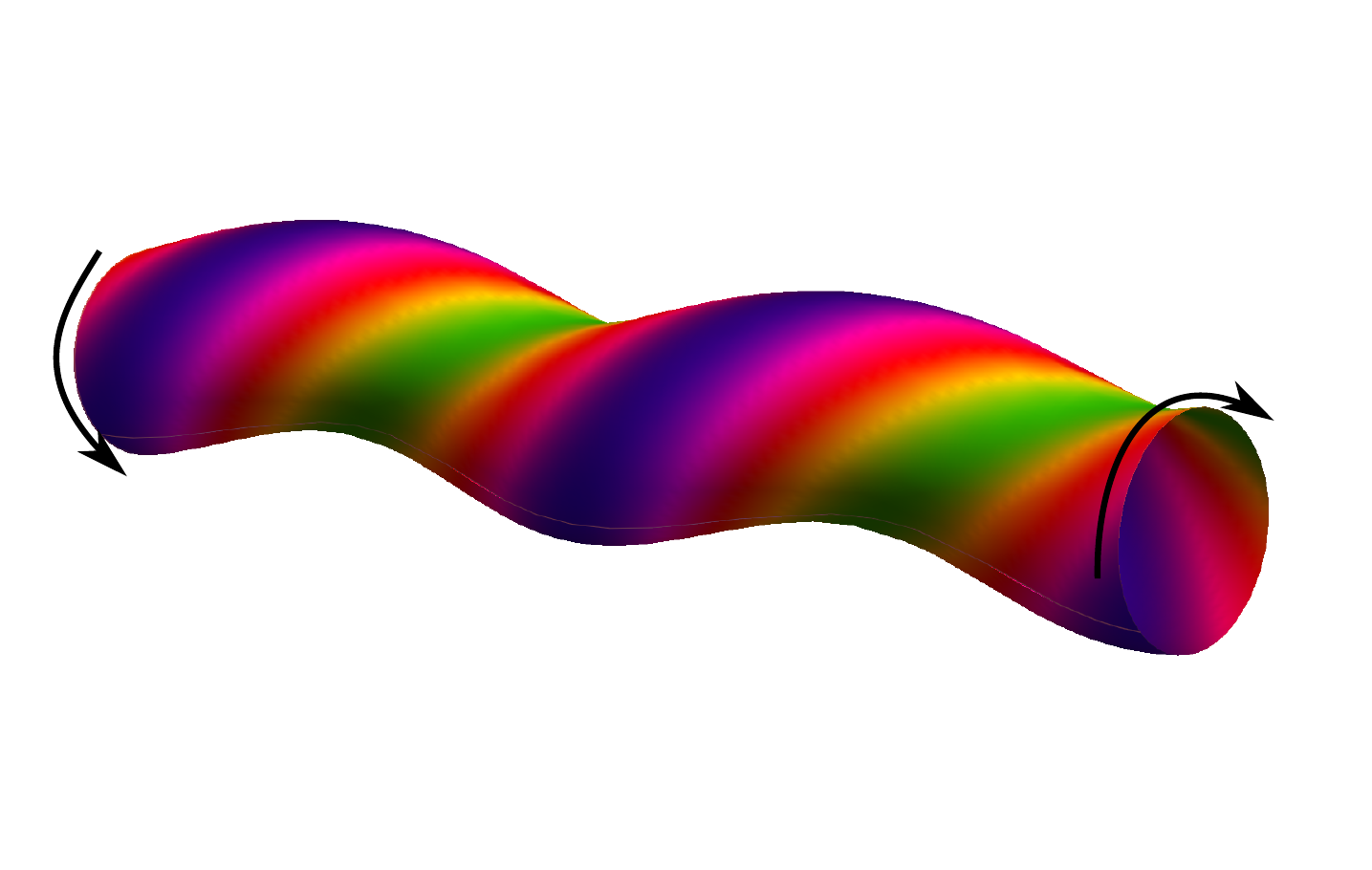}
\caption{\label{fig:normalForces}The normal component of the viscous force per unit area on a helical shaped tube coming from the term $\eta_m b^{i}{}_j\nabla_i v^j$ in Eq.~\ref{eq:shape} (purple outward, green inward). This shows the helical nature of the instability above the critical shear rate. The tube would eventually be advected to stability again, as the helix winds up under the shear flow (see text), although the tube may reach the nonlinear deformation regime before this happens.}
\end{figure}

Note the sign of the final term in the $\mathcal{F}^u_{q,m}$ coefficient, scaling with the shear $\Omega$, suggests that the shear flow could lead to an instability in the $m\neq 0$ modes, see Fig.~\ref{fig:normalForces}. Note that the ($m\to-m$, $\tilde{q}\to-\tilde{q}$) symmetry of the normal force defines a ``handedness'' which changes upon reversing the direction of the shear rate.

Similar linear response equations can be found for the force balance and continuity in the tangential directions, these can then be used to solve for $\bar{\delta v}^{z}_{q,m}$, $\bar{\delta v}^{\theta}_{q,m}$ and $\bar{\delta\sigma}_{q,m}$ in terms of $\bar{u}_{q,m}$ and its time derivative. From this we derive the following growth rate equation for $\bar{u}_{q,m}$, where time is normalised according to $t=\tilde{t}\tau$ with $\tau=\frac{\eta_m}{\sigma_0}$,
\begin{equation}\label{eq:fullGrowthRate}
\partial_{\tilde{t}}\bar{u}_{q,m} = -\dot\imath m\frac{v_0\eta_m}{r_0\sigma_0}\bar{u}_{q,m} -\tilde{\Omega}m\partial_{\tilde{q}}\bar{u}_{q,m}+ F(q,m)\bar{u}_{q,m}
\end{equation}
where
\begin{equation}\label{eq:stabilityFunction}
\begin{split}
&F(q,m) = \Big[m\tilde{q}\left(\left(m^2+\tilde{q}^2 \right)^2-2 \tilde{q}^2\right)\tilde{\Omega} -\\ &\left(m^2+\tilde{q}^2\right)^2 \left(1+m^4 +\tilde{q}^4 +2 m^2\left(\tilde{q}^2-1\right)\right)\Big]\left(2\tilde{q}^4\right)^{-1}
\end{split}
\end{equation}
and $\tilde{\Omega}=\frac{\eta_m \Omega}{\sigma_0}$ is the dimensionless shear.

The modes become unstable when the real part of the growth rate changes sign to $\Re\left\{F(m,q)\right\}>0$, which occurs for
\begin{equation}\label{eq:stabilityCriterion}
\tilde{\Omega} m \tilde{q} >\frac{\left(m^2+\tilde{q}^2\right)^2\left(1+m^4 +\tilde{q}^4+2 m^2\left(\tilde{q}^2-1\right)\right)}{\left(m^2+\tilde{q}^2\right)^2-2\tilde{q}^2}\text{.}
\end{equation}

\begin{figure}[b]
\includegraphics[width=0.45\textwidth]{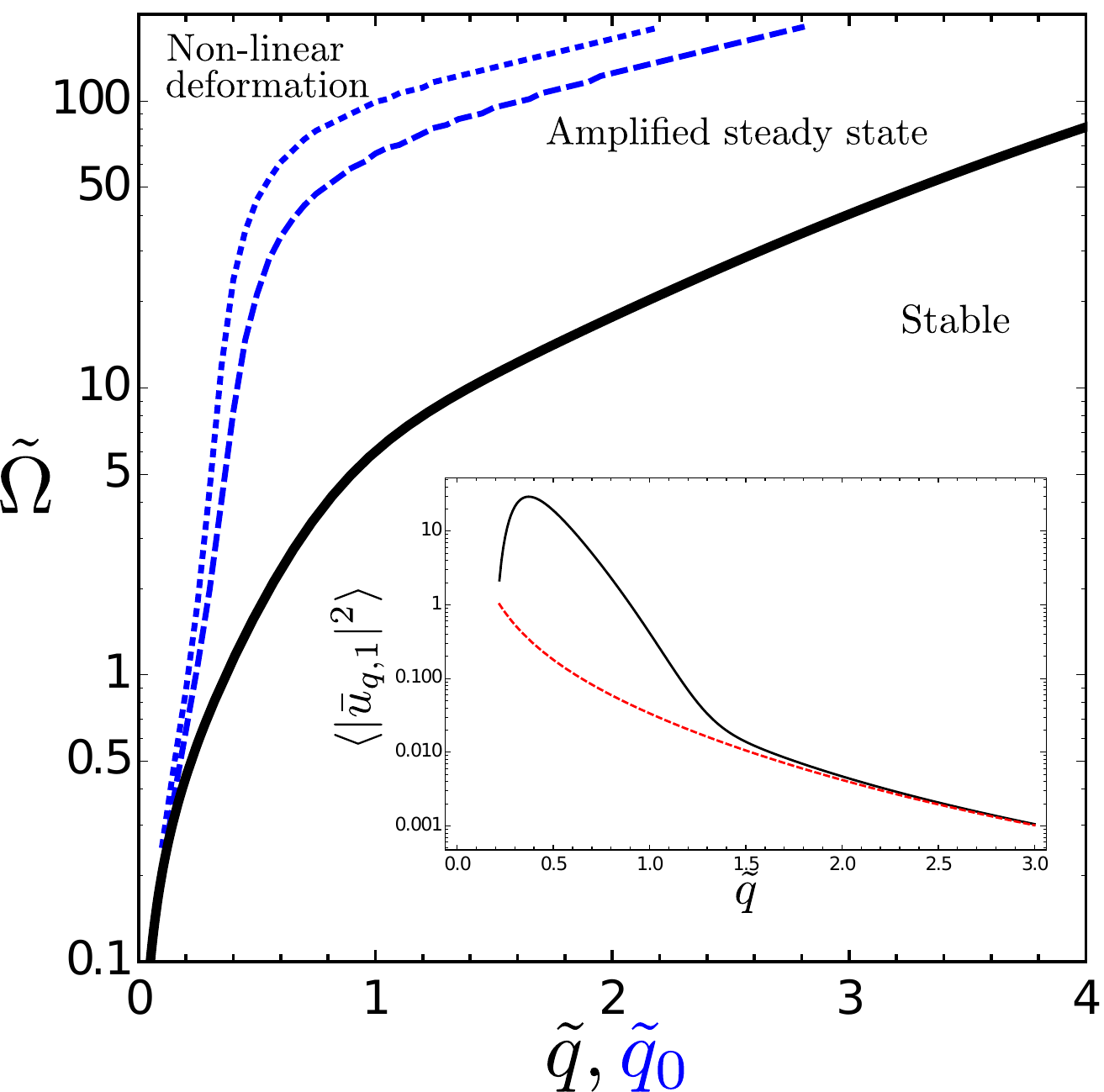}
\caption{\label{fig:growthAndStability}Dynamical phase diagram of the state of the tube in the presence of fluctuations. Helical $\tilde{q}$-modes to the left of the black line are unstable according to Eq.~\ref{eq:stabilityCriterion}. The blue dashed and blue dotted lines show the value of $\tilde{\Omega}$ for which $5\%$ and $32\%$ of the fluctuations become non-linear as a function of $\tilde{q}_0$, the cutoff on the noise spectrum coming from the finite tube length. Inset shows the statistical steady state of $\langle |\bar{u}_{q,1}|^2\rangle$ for cut-off $\tilde{q}_0=0.2$ and shear $\tilde{\Omega}=1$ in black with equilibrium thermal fluctuations shown in dashed red.}
\end{figure}

The $m=0$ peristaltic mode is always linearly stable. This is not the case for the $m=1$ mode, which is the first to be driven unstable. The stability threshold for the $m=1$ mode is plotted in Fig.~\ref{fig:growthAndStability} in solid black. Note that the growth rate is a discrete function of $\tilde{q}=\frac{2\pi n r_0}{L}$ with discretization set by the length of the tube. This means that, beyond a certain rotation speed, a helical mode will grow, with pitch length initially set by the length of the tube. The apparent divergence of the growth rate for small $\tilde{q}$ is ultimately limited by bulk hydrodynamics.

This analysis is complicated by the advection in $\tilde{q}$ space of helical modes that arises from the term involving $\partial_{\tilde{q}}$ in Eq.~\ref{eq:fullGrowthRate}. This reflects the fact that the ground-state shear flow continuously adds new turns to an existing helix, thereby increasing its characteristic wave number. Large wave numbers are stable, so a helical perturbation rendered unstable by the shear flow is eventually stabilised by this advection. 
This leads to a non-equilibrium steady state, which can be obtained by solving Eq.~\ref{eq:fullGrowthRate} with thermal noise added using the method of stochastic characteristics, see S.I.~and \cite{chow_stochastic_2014} for details. This non-equilibrium steady state for $\langle |\bar{u}_{q,m}|^2\rangle$ has a peak in $q$-space, see Fig.~\ref{fig:growthAndStability} inset. Because the $m=1$ modes are critical in the $\tilde{q}\to 0$ limit \cite{fournier_critical_2007} we choose a small $\tilde{q}$ cut-off for the noise spectrum at $\tilde{q}_0=2\pi r_0/L$, which is physical, given the finite length of our tube. Eq.~\ref{eq:fullGrowthRate} is based on a small perturbation expansion and breaks down when the spatial gradients become large, $\langle|\nabla u|^2\rangle\sim 1$, in which case the end state might be quite different. In Fig.~\ref{fig:growthAndStability}, we plot the value of $\tilde{\Omega}$, as a function of the cutoff, $\tilde{q}_0$, for which $5\%$ and $32\%$ of fluctuations are in the non-linear regime as the blue dashed and blue dotted lines. Beyond this shear rate the tube will be deformed non-linearly and it is not clear if there will be a steady state. A full analysis of this is beyond the scope of the present work.

In the small $\tilde{q}$ limit, the threshold shear (Eq.~\ref{eq:stabilityCriterion}) is $\tilde{\Omega}\approx 2\tilde{q}$ (see S.I.). The shear rate is $\Omega\sim \frac{2\pi r_0\nu}{L}$, where $\nu$ is the spinning frequency. Assuming that the cut-off wavenumber of the tube is associated with a fundamental mode  $\tilde{q}_{0}=\frac{2\pi r_0}{L}$, gives the critical spinning frequency for the onset of instability as $\nu_{\text{crit}}\simeq \frac{2\sigma_0}{\eta_m}$. The functional form of the critical frequency can be understood using a scaling analysis of Eq.~\ref{eq:shape}, see S.I..

Typical membranes in the fluid (liquid disordered) phase have viscosities \mbox{$\eta_m\sim 10^{-9}-10^{-8}\>\text{Pa}\>\text{m}\>\text{s}$} \cite{hormel_measuring_2014} (higher in the liquid ordered phase). However, much higher values of effective viscosity have been associated with tubes pulled from living cells, $\eta_m\sim 10^{-7}-10^{-5}\>\text{Pa}\>\text{m}\>\text{s}$ \cite{brochard-wyart_hydrodynamic_2006}. We use these numbers, noting that effective viscosities may be higher still if the neck is crowded with proteins. We assume the surface tension takes a physiologically typical value \footnote{Noting that this value may vary enormously near the neck of a budding vesicle subject to forces, e.g. from neighbouring actomyosin.} of $\sigma_0\sim 10^{-5}\>\text{N}\>\text{m}^{-1}$ \cite{roux_reaching_2014,antonny_membrane_2016}.
Vesicular necks correspond to short tubes with $\tilde{q}_0\sim 1$ so, from Fig.~\ref{fig:growthAndStability} we find 
$\tilde{\Omega}\sim 5$ for the stability criterion and $\tilde{\Omega}\sim 50$ for the non-linearity criterion which correspond to $\nu\sim5-500\text{Hz}$ and $\nu\sim50-5000\text{Hz}$ respectively, with the wide range traced to the uncertainty in membrane viscosity. Dynamin polymers have been measured to have rotational frequencies $\nu\sim 10\text{Hz}$ \cite{roux_gtp-dependent_2006}, suggesting the instability could be accessible to dynamin for the higher values of viscosity found in cells. These estimates are quite conservative as in a realistic scenario active fluctuations are likely to be much larger than thermal fluctuations, perhaps by an order of magnitude or more, and we are unlikely to have such a hard cut-off at $\tilde{q}_0$.

A natural way for the fluctuations to progress in the non-linear regime is fission of the tube, which is of particular significance given that the exact mechanism for dynamin mediated fission is unknown. As the fluctuations grow the surface tension will increase, either narrowing the tube or causing Pearling \citep{nelson_dynamical_1995}. An increase in tension has been shown to accelerating spontaneous tube fission \citep{morlot_membrane_2012} and friction impeding membrane flow has been shown experimentally to scission tubes \citep{simunovic_friction_2017}. The increase in fluctuations is also likely to promote the formation of hemi-fused states, which can be an important intermediate for fission \citep{pannuzzo_role_2018}. Surface tension fluctuations, even at the linear level, can be estimated to be much larger than the ground-state surface tension and this could also be important in driving membrane lysis, see S.I.. This picture of fission, promoted by membrane hydrodynamics just outside the active dynamin site, is consistent with the experimental observation that the location of fission is near the edge of the active dynamin site rather than directly under it \citep{morlot_membrane_2012}. The time-scale over which the instability grows is of the order of $\tau\sim 10^{-2}-1\>\text{s}$, which is sufficiently fast to be consistent with the dynamin-induced fission process \citep{dar_high-throughput_2015}.

Although we have provided evidence that a membrane instability can be driven by the rotation of dynamin, our study is based on the simplified geometry of a cylindrical tube, rather than the neck of a budding vesicle, a location where dynamin might typically act \textit{in-vivo}. While our approach becomes analytically intractable for such complex membrane geometries we can gain some intuition into how the driving force per unit area of the instability changes with the geometry of the neck region by considering the term in the normal force balance equation that is responsible for driving the instability. Given the helical symmetry of the instability we infer that this driving force-per-unit-area goes like the mixed derivative in the shape, $f_{\text{driving}}\sim \eta_m b^i{}_j\nabla_i v^j$. The term which acts like the shear rate on the tube now depends on $z$ and we must calculate it numerically, see S.I.. In the case of a catenoidal neck this leads to an amplification of the driving force by (only) a factor of $2$ near the active site ($z=0$), for details see S.I.. Whilst a relatively small effect, this is qualitatively consistent with the experimental observation that dynamin fission of a tube \textit{in-vitro} often occurs near the GUV neck \citep{morlot_membrane_2012} and that fission on the necks of budding vesicles \textit{in-vivo} occurs faster than it does on long tubes \citep{morlot_deformation_2010,roux_reaching_2014}. 

A second possibility for the non-linear growth is a stable non-equilibrium shape driven by the membrane flow. In this case it is worth noting an analogy between the membrane tube instability discussed here and elastic rods under torsion that deform nonlinearly into plectonemes \citep{audoly_elasticity_2010}. If excess membrane area is more readily available it may be possible for the unstable tube to develop {\it fluid} plectonemes if the instability develops without a scission-inducing increase in tension. Similar structures are  observed in experiments on long tubes covered in dynamin \citep{roux_gtp-dependent_2006,morlot_deformation_2010}.

The experiment suggested in Fig.~\ref{fig:diagram}b would both test  our predictions more quantitatively and probe the non-linear evolution of the fluctuations so as to determine whether these hydrodynamic effects alone are sufficient to induce fission. The instability should also arise in a longer tube, however the quantitative nature of our predictions would likely require modifications due to screening of membrane flow by the ambient fluid. In this case we expect that the unstable wavelength would then be set by the screening length $\sqrt{L_{\text{SD}}r_0}$ rather than the tube length \citep{henle_hydrodynamics_2010,ferziger_computational_2002} and that our results would continue to hold at the scaling level.

In summary, we have developed a hydrodynamic theory that predicts an instability on fluid membrane tubes that is driven purely by a shear in the membrane flow. Such flows are shown to first drive a helical instability, which is quite distinct from any previously identified instabilities of fluid membrane tubes. This instability, although eventually advected to stability by the flow is shown to be able to significantly modify and enhance the fluctuation spectra of a membrane tube. We predict that this instability, and perhaps its fully nonlinear manifestation, may be physiologically accessible to dynamin. Such hydrodynamic effects have not previously been considered in models of its function \citep{lenz_mechanochemical_2008,morlot_deformation_2010}. This instability may provide a mechanism for  dynamin-mediated tube scission.

\acknowledgments{The authors acknowledge very helpful comments from S.~Ramaswamy (Bangalore) and helpful discussions with P.~Bassereau (Institut Curie), R.~Phillips (CalTech), and J.~E.~Sprittles, G.~Rowlands and J.~Binysh (Warwick). SCAI~would like to acknowledge funding from the UK EPSRC under grant number EP/L015374/1, the Centre for Doctoral Training in Mathematics for Real-World Systems and support from the Labex CelTisPhyBio (ANR-11-LABX-0038, ANR-10-IDEX-0001-02).}

\bibliography{bibliography}

\clearpage
\onecolumngrid
\makeatletter 
\def\tagform@#1{\maketag@@@{(S\ignorespaces#1\unskip\@@italiccorr)}}
\makeatother
\makeatletter
\makeatletter \renewcommand{\fnum@figure}
{\figurename~S\thefigure}
\makeatother
\def\eq#1{{Eq.(S\ref{#1})}}    \def\fig#1{{Fig.S\ref{#1}}}
\setcounter{figure}{0} 
\setcounter{equation}{0} 
\appendix
\section*{Supplementary Information}\label{sec:SI}
\subsection*{Differential Geometry}
Here we present a ``users guide'' to the style of geometric notation used in the main paper. We do not focus on mathematical rigour here, for a more formal treatment see \cite{frankel_2011}.

If we define a manifold $\mathcal{M}^n$ where the derivative of a curve at point $p\in \mathcal{M}^n$ gives an element of the tangent space $\boldsymbol{X}_p\in\mathcal{T}_p\left(\mathcal{M}^n\right)$, we can express this in terms of a coordinate basis
\begin{equation}
\boldsymbol{X}_p = X^i \left(\frac{\partial}{\partial x^i}\right)_p = X^i \left(\vec{e}_i\right)_p
\end{equation}
where Einstein summation over mixed indices is implicit.

If we choose a family of curves on $\mathcal{M}^n$ with continuous derivatives we can extend the definition of the tangent space to the tangent bundle on $\mathcal{M}^n$, $\mathcal{T}\left(\mathcal{M}^n\right)= \cup_p \mathcal{T}_p\left(\mathcal{M}^n\right)$. This extends the definition of a vector to a vector field on the the manifold, $X\in \mathcal{T}\left(\mathcal{M}^n\right)$.

The dual of $\mathcal{T}\left(\mathcal{M}^n\right)$ can be defined as the cotangent space $\mathcal{T}^*\left(\mathcal{M}^n\right)$. An element of this space, a 1-form, is defined in the following way $\boldsymbol{\omega}\in\mathcal{T}^*\left(\mathcal{M}^n\right)$
\begin{equation}
\boldsymbol{\omega}\left(X\right) \to \mathbb{R}\text{.}
\end{equation}

In coordinate notation
\begin{equation}
\boldsymbol{\omega}\left(\boldsymbol{X}\right) = \omega_i X^j \mathrm{d}x^i \frac{\partial}{\partial x^j} = \omega_i X^j \delta^i_j = \omega_i X^i\text{.}
\end{equation}

In general a type $(p,q)$ tensor field, $T$ is defined in the following way
\begin{equation}
T\left(X_1,...,X_p,\omega_1,...,\omega_q\right)\to\mathbb{R}
\end{equation}
where $X_1,...,X_p\in \mathcal{T}\left(\mathcal{M}^n\right)$ and $\omega_1,...,\omega_q\in \mathcal{T}^*\left(\mathcal{M}^n\right)$.

We can define a type $(2,0)$ metric tensor on the manifold as
\begin{equation}
g(\cdot ,\cdot) :\quad g\left(X,Y\right)\to \mathbb{R}
\end{equation}
where $X\text{,}Y\in \mathcal{T}\left(\mathcal{M}^n\right)$. 
\begin{equation}
g(\cdot,\cdot)=\text{d}s^2 = g_{ij}\mathrm{d}x^i\mathrm{d}x^j = \vec{e}_i\cdot \vec{e}_j \mathrm{d}x^i\mathrm{d}x^j
\end{equation}
which allows a mapping between vectors and 1-forms.

The exterior or wedge product between two $1$-forms is defined as the totally asymmetric tensor product
\begin{equation}
\boldsymbol{\omega}_1 \wedge \boldsymbol{\omega}_2 = \boldsymbol{\omega}_1 \otimes \boldsymbol{\omega}_2 - \boldsymbol{\omega}_2 \otimes \boldsymbol{\omega}_1\text{.} 
\end{equation}

A $p$-form, $\alpha$, can be defined from $p$ $1$-forms as 
\begin{equation}
\alpha = \boldsymbol{\omega_1}\wedge ...\wedge \boldsymbol{\omega}_p\text{.}
\end{equation}

This has the following property
\begin{equation}
\boldsymbol{\omega}_1\wedge ...\wedge\boldsymbol{\omega}_r\wedge ... \wedge\boldsymbol{\omega}_s \wedge ... \boldsymbol{\omega}_p = - \boldsymbol{\omega}_1\wedge ...\wedge\boldsymbol{\omega}_s\wedge ... \wedge\boldsymbol{\omega}_r \wedge ... \boldsymbol{\omega}_p
\end{equation} 
for any two $s$, $r$. Or in coordinate notation
\begin{equation}
a_{i...r...s...j}=-\alpha_{i...s...r...j}
\end{equation}
where $\alpha=\alpha_{i...j}\mathrm{d}x^i\wedge ...\wedge\mathrm{d}x^{j}$.

This along with the metric leads to the natural geometric definition of the volume form $\text{vol}^n :=\sqrt{g}\mathrm{d}x^1\wedge ... \wedge\mathrm{d}x^n$, where $g:=\text{det}(g_{ij})$.

The exterior derivative, $\boldsymbol{d}$, of a smooth function $f$ is just its differential $\boldsymbol{d}f=\frac{\partial f}{\partial x^i}\mathrm{d}x^i$.
The exterior derivative, $\boldsymbol{d}$, of a $p$ form is a $p+1$ form
\begin{equation}
\boldsymbol{d}\alpha = \boldsymbol{d}\alpha_{i...j}\wedge\mathrm{d}x^i\wedge ...\wedge\mathrm{d}x^{j}\text{.}
\end{equation}

The Hodge star operator, $\star:\tau^*(\mathcal{M})^{(k)}\to \tau^*(\mathcal{M})^{(n-k)}$, is defined by the Hodge inner product of two differential forms $\alpha$ and $\beta$
\begin{equation}
\alpha\wedge\star\beta = \left(\alpha\cdot\beta\right)\text{vol}^{n}
\end{equation}
in coordinate notation we have
\begin{equation}
\star\alpha = \epsilon_{i_1\dots i_n}\sqrt{\det{g}}\alpha_{j_1\dots j_k} g^{i_1j_1}\dots g^{i_kj_k}\mathrm{d}x^{i_{k+1}}\wedge\dots\wedge \mathrm{d}x^{i_n}
\end{equation}
where $\epsilon$ is the totally asymmetric tensor.

A diffeomorphism is a map between two manifolds that is smooth, one-to-one, onto and has a smooth inverse. The Lie derivative is a natural object to use in continuum mechanics as it describes how a vector field $Y$ changes along the flow generated by a vector field $X$. If $\phi(t)=\phi_t$ is a diffeomorphism parametrised by $t$ and describing the local flow generated by $X$, where $t$ is defined such that $\lim_{t\to 0 }\phi_t(X)=X$, then we define the Lie derivative of a vector field $Y$ with respect to a vector field $X$ as follows
\begin{equation}
[\mathcal{L}_X Y]_x =\lim_{t\to 0} \frac{[\phi_{-t*}Y_{\phi_tx} - Y_x] }{t} = X(Y)-Y(X)
\end{equation}
as such $\mathcal{L}_X Y$ is a vector field on $\mathcal{M}^n$. Similar identities can be derived for more general tensors \cite{frankel_2011}.

We will define the Laplace-Beltrami operator as
\begin{equation}
\Delta_{\text{LB}}=-\star\boldsymbol{d}\star\boldsymbol{d}
\end{equation}
which for scalar $\phi$ and vector $\boldsymbol{v}$ is the following in index notation
\begin{equation}
\begin{split}
&\Delta_{\text{LB}}\phi = -\frac{1}{\sqrt{|g|}}\partial_i\left(\sqrt{|g|}g^{ij}\partial_j\phi\right)\\
&\Delta_{\text{LB}}v^{q} = -\sqrt{|g|}\epsilon_{np}\epsilon_{kl}g^{pq}g^{nm}\partial_m\left(\sqrt{|g|}g^{kj}g^{li}\partial_j\left(v^rg_{ri}\right)\right)
\end{split}
\end{equation}
where the later formula is not usually given in the literature as it is simpler to work with exterior calculus identities (which is how we will proceed).

One final point of note is that we will use the $\flat$, $\sharp$ notation to denote raising and lowering of indices for conciseness. For example, if $\boldsymbol{v}\in \mathcal{T}\left(\mathcal{M}^n\right)$ and $\omega\in \mathcal{T}^*\left(\mathcal{M}^n\right)$, then
\begin{equation}
\begin{split}
&\boldsymbol{v}^\flat=g_{ij}v^{j}\mathrm{d}x^{i} = v_i\mathrm{d}x^{i}\\
&\omega^\sharp = g^{ij}\omega_{j}\vec{e}_{i} = \omega^{i}\vec{e}_{i} \text{.}
\end{split}
\end{equation}

\subsection*{Hydrodynamics on moving fluid membranes}
We need to construct force balance and mass conservation equations on a moving membrane which we will denote by Riemannian manifold $\Gamma$. As $\Gamma$ will be embedded in $\mathbb{R}^3$ we denote vector fields living in $\mathbb{R}^3$ with an arrow above them, for example $\vec{x}$, and vector fields living in the tangent bundle of $\Gamma$ by bold typeface, e.g.~$\boldsymbol{x}$.

The position of $\Gamma$ will be denoted by $\vec{X}_\Gamma(x_1,x_2)$, which depends local on two coordinates of $\mathbb{R}^3$. This allows for the definition of a basis on $\Gamma$, $\vec{e}_i=\partial_i\vec{X}$. $\Gamma$ is equipped with a metric $\text{d}s^2=g_{ij}\text{d}x^i \text{d}x^j$, where $g_{ij}=\vec{e}_i\cdot\vec{e}_j$, this and it's inverse act to raise and lower indices respectively (the action by the metric of raising and lower of indices will sometimes be denoted by the $\sharp$ and $\flat$ signs respectively). The triad $(\vec{e}_1,\vec{e}_2, \vec{n}=\frac{\vec{e}_1\times\vec{e}_2}{|\vec{e}_1\times\vec{e}_2|})$ forms a local frame on $\Gamma$. We also denote the second fundamental form on $\Gamma$ as $\text{d}B = b_{ij}\text{d}x^i\text{d}x^j$ where $b_{ij}=\vec{n}\cdot \left(\partial_{j}\vec{e}_i\right)$. The connections along the tangent and normal bundles are defined in the following way
\begin{equation}
\partial_i\vec{e}_j = C^k{}_{ij}\vec{e}_k;\quad \partial_i\vec{n}=-b_i{}^j\vec{e_j}
\end{equation}
where $C^i{}_{jk}=\frac{1}{2}g^{im}\left( \partial_jg_{mk} +\partial_kg_{jm} -\partial_mg_{jk}\right)$ are Christoffel symbols. We will also define the mean curvature, $H$, and Gaussian curvature, $K$, in the following manner
\begin{equation}
2H=b_i{}^i;\quad K=\text{det}\left(b_i{}^j\right)\text{.}
\end{equation}

\subsubsection{Flows on moving curved surfaces}
Formally, the rate-of-deformation tensor for a manifold is defined as the Lie-Derivative of the metric along the velocity field ($\vec{V}=\boldsymbol{v}+ w\vec{n}$), this can be shown to be equal to \cite{marsden_mathematical_1994,arroyo_relaxation_2009}
\begin{equation}
d=\frac{1}{2}\mathcal{L}_{\vec{V}}\left( g\right)= \frac{1}{2}\left( \boldsymbol{\nabla}\boldsymbol{v}^\flat + \left(\boldsymbol{\nabla}\boldsymbol{v}^\flat\right)^T\right) - b w
\end{equation}
where $\boldsymbol{\nabla}$ is the covariant derivative. The first two terms are covariant versions of the standard rate-of-deformation tensor, whereas the third term describes the coupling between curvature, $b$, and the velocity normal to the membrane, $w$.

A simple heuristic derivation of this can be obtained using simple local constructions. If we consider a membrane which when un-deformed, $\mathcal{M}$, and is approximately flat then its line element (metric) can be written
\begin{equation}
\mathrm{d}s^2=\mathrm{d}x^2+\mathrm{d}y^2\text{.}
\end{equation}
If we deform this manifold by the vector $\left(\phi_x,\phi_y,\psi\right)$ to a new manifold $\mathcal{M}'$ and choose coordinates $x$, $y$ such that the second fundamental form of of $\mathcal{M}'$ is given by
\begin{equation}
b=\left(\begin{matrix}
\mathrm{d}x & \mathrm{d}y\end{matrix}\right)\left(\begin{matrix}
k_1 & 0\\
0 & k_2
\end{matrix}\right)\left(\begin{matrix}
\mathrm{d}x\\
\mathrm{d}y\end{matrix}\right)\text{.}
\end{equation}

The new metric on the surface $\mathcal{M}'$ is given by $\mathrm{d}s'^2 = (\mathrm{d}x')^2+(\mathrm{d}y')^2$ where, to lowest order,
\begin{equation}
\begin{split}
\mathrm{d}x' = \left(1-k_1\psi\right)\left(1+\partial_x\phi_x\right)\mathrm{d}x + \partial_y\phi_x\mathrm{d}y\\
\mathrm{d}y' = \left(1-k_2\psi\right)\left(1+\partial_y\phi_y\right)\mathrm{d}y + \partial_x\phi_y\mathrm{d}x\text{.}
\end{split}
\end{equation}
so the new metric is given by
\begin{equation}
\begin{split}
\mathrm{d}s'^2 = \left(1-2 k_1\psi -2 \partial_x\phi_x\right)\mathrm{d}x^2 + 2\left(\partial_x\phi_y+\partial_y\phi_x\right)\mathrm{d}x\mathrm{d}y\\ + \left(1-2 k_2\psi -2 \partial_y\phi_y\right)\mathrm{d}y^2
\end{split}
\end{equation}
up to linear order in the variables $(\phi_x,\phi_y,\psi)$ and their derivatives.

If we assume $\left(\phi_x,\phi_y,\psi\right) = \Delta t\left(v_x,v_y,w\right)=\Delta t \vec{V}$, where $\vec{V}$ is the membrane velocity then we can write the Lagrangian deformation tensor, $L$, as
\begin{equation}
\begin{split}
L&=\frac{1}{2}\left[\mathrm{d}s'^2-\mathrm{d}s^2\right]\\ &= \Delta t\left[\left(\partial_xv_x-k_1w\right)\mathrm{d}x^2 + \left(\partial_xv_y+\partial_yv_x\right)\mathrm{d}x\mathrm{d}y +  \left(\partial_yv_y-k_2w\right)\mathrm{d}y^2\right]
\end{split}
\end{equation}
and dividing by $\Delta t$ and taking the limit $\Delta t\to 0$ gives the rate of deformation tensor 
\begin{equation}
d = \left[\frac{1}{2}\left(\nabla_iv_j+\nabla_jv_i\right) - w b_{ij}\right]\mathrm{d}x^i\otimes\mathrm{d}x^j\text{.}
\end{equation}

We can find the continuity equation (incompressibility condition) for the membrane by taking the trace of the rate-of-deformation tensor, $d$,
\begin{equation}
\boldsymbol{\nabla}\cdot\boldsymbol{v}=2H w\text{.}
\end{equation}

\subsubsection{Curvature energies}
The membrane also has associated curvature energies given by the Helfrich energy
\begin{equation}
E_{\text{Hel}}=\int_\Gamma \mathrm{d}A_{\Gamma}2\kappa H^2
\end{equation}
the time derivative of which depends only on $w$, $\partial_tE_{\text{Hel}}=\dot{E}[w]$ \cite{rahimi_curved_2013}. Defining the Rayleigh dissipation functional for the membrane in the following way
\begin{equation}
\quad W_\Gamma= \int_{\Gamma}\eta_m d:d \: \mathrm{d}A_\Gamma
\end{equation}
accounts for the fluid behaviour of the membrane. From this a complete dissipation functional for the system can be defined as
\begin{equation}\label{eq:energyFunctional}
\begin{split}
G=W_\Gamma +\dot{E}+\int_\Gamma\sigma\left(\boldsymbol{\nabla}\cdot\boldsymbol{v}-2H w\right)\mathrm{d}A_\Gamma 
\end{split}
\end{equation}
imposing incompressibility of membrane with Lagrange multiplier, $\sigma$, which corresponds to surface tension. Performing functional variation with respect to the components of the surface velocity yields the force balance equations in the main text, see \cite{arroyo_relaxation_2009} for details.

\subsection*{Ground-state flows}
We consider a problem of a membrane tube with  spinning velocity $v_0$ at $z=0$, attached to a flat membrane at $z=L$ where $L\ll L_{\text{SD}}$ such that we can solve for the ground-state using only the membrane equations. We treat this flat membrane as an effective ``impedance'' acting at the end of the tube, as such we do not balance the shape equations at $z=L$. 

We may want to consider a tube attached to a sheet of membrane that has some friction associated to some underlying molecular interactions. For example, consider that the tube has been pulled from the plasma membrane which is attached to the acto-myosin network \cite{kaksonen_mechanisms_2018}. We model this using D'arcy's equation on the sheet
\begin{equation}
\frac{1}{r}\partial_r\left(r\partial_rv\right)-\frac{v}{r^2}-\frac{\lambda}{\eta_m} v=0
\end{equation}
where $\lambda$ is a friction coefficient associated with the adhesions.
The solution to this equation is of the form $v=AK_1\left(\sqrt{\frac{\lambda}{\eta_m}}r\right)$, where $K_i(x)$ is a modified Bessel equation of the second kind of order $i$. We solve both geometries for some velocity $v_L$ and then balance torques to find the ground-state velocity of the tube.

This leads a velocity profile on the tube (where the flow just follows the standard Stokes equations) of the form
\begin{equation}\label{eq:groundStateSI}
\boldsymbol{v}=\left(v_0 - \Omega z\right)\vec{e}_\theta
\end{equation} where $\Omega=\frac{v_0\sqrt{\frac{\lambda}{\eta_m}}\frac{K_2}{K_1}}{1+L\sqrt{\frac{\lambda}{\eta_m}}\frac{K_2}{K_1}}$ where $K_i=K_i\left(\sqrt{\frac{\lambda}{\eta_m}}r_0\right)$.

In the limit $\lambda\to 0$ we recover the solution with no friction, where $\Omega=\frac{2 v_0}{2L+r_0}$.

In both of this and the $\lambda\to\infty$ limit the shear rate is of a similar order of magnitude, scaling like $\Omega\sim v_0/L$.

\subsection*{Geometry and flows on tubes with small deformations}
We now consider a perturbation to the geometry of the tube of the form $r(\theta,z,t)=r_0 + u(\theta,z,t)$. We will assume that this perturbation is small with respect to the radius, $u/r_0\ll 1$. We take the normal to be outward in the radial direction, and project forces in the normal along this axis. All components of differential forms are given in the basis $\mathrm{d}\theta$, $\mathrm{d}z$ hence the different dimensions in components.

To linear order the metric and its inverse on the membrane are
\begin{equation}
[g_{ij}] = \begin{bmatrix}
r_0^2 +2 r_0 u & 0\\
0 & 1
\end{bmatrix}; \quad \quad g^{-1}= [g^{ij}] = \begin{bmatrix}
\frac{1}{r_0^2} -\frac{2u}{r_0^3} & 0\\
0 & 1
\end{bmatrix}
\end{equation}
The second fundamental form (and its mixed index version) are given by the following at linear order
\begin{equation}
[b_{ij}] = \begin{bmatrix}
\partial_\theta^2u -  r_0 - u & \partial_{z\theta}u\\
\partial_{z\theta}u & \partial_z^2u
\end{bmatrix};\quad
[b_{i}{}^{j}] = \begin{bmatrix}
\frac{\partial_\theta^2u}{r_0^2} -  \frac{1}{r_0} - \frac{u}{r_0^2} & \frac{\partial_{z\theta}u}{r_0^2}\\
\partial_{z\theta}u & \partial_z^2u
\end{bmatrix};
\end{equation}
which gives mean and Gaussian curvature 
\begin{equation}
\begin{split}
&2H = b_{i}{}^{i}= b_{ij}g^{ji}= \frac{\partial_\theta^2u}{r_0^2} - \frac{1}{r_0} + \frac{u}{r_0^2} +\partial_z^2u\\
&K = \det\left(b_i^j\right)=\det\left(b_{ik}g^{kj}\right)= -\frac{\partial_z^2u}{r_0}
\end{split}\text{.}
\end{equation}

The Christoffel symbols are the following
\begin{equation}
C^\theta_{\text{  }ij} = \begin{bmatrix}
\frac{\partial_\theta u}{r_0} & \frac{\partial_z u}{r_0}\\
\frac{\partial_zu}{r_0} & 0
\end{bmatrix}; \quad \quad C^z_{\text{  }ij} = \begin{bmatrix}
-r_0\partial_zu & 0\\
0 & 0
\end{bmatrix}
\end{equation}
which can be used to find the covariant derivative of the velocity field on the membrane $\boldsymbol{v} = (v +\delta v^\theta)\vec{e}_\theta +\delta v^z \vec{e}_z$
\begin{equation}
\boldsymbol{\nabla}\boldsymbol{v} = \begin{bmatrix}
\frac{1}{r_0}\partial_\theta\delta v^\theta  & \partial_\theta\delta v^z\\
-\frac{\Omega}{r_0}+\frac{1}{r_0}\partial_z\delta v ^\theta  & \partial_z\delta v^z
\end{bmatrix}\text{.}
\end{equation}

We will make use of this to calculate the viscous part of the normal membrane response in the shape equation
\begin{equation}
b^{\sharp}:\boldsymbol{\nabla}\boldsymbol{v} = -\frac{1}{r_0^2}\partial_\theta\delta v^\theta  -\frac{\Omega}{r_0}\partial_{z\theta}u\text{.}
\end{equation}

We also note here the Hodge duals of the fundamental forms as this provides a natural way to compute Laplacians on manifolds
\begin{equation}
\begin{split}
&\star \text{vol}^2 = 1;  \quad\quad\quad\quad \quad \quad\star 1 =\text{vol}^2\\
&\star\text{d}\theta = \left(\frac{1}{r_0} - \frac{u}{r_0^2}\right)\text{d}z  \quad\star \text{d}z = -(r_0 +u)\text{d}\theta
\end{split}
\end{equation}
we find the Laplacian of the mean curvature $-\star\boldsymbol{d}\star\boldsymbol{d} H$ in order to derive the bending rigidity dominated response. After some lengthy algebra and taking the Fourier representation $u=\sum_{q,m}\bar{u}_{q,m} e^{\dot{\imath} qz +\dot{\imath} m\theta}$ with similar transforms for $\sigma=\sigma_0+\delta\sigma$ and the surface velocity components, we can write the shape equation as a linear response theory. This gives Eq.~\ref{eq:linearResponseShapeTerm} in the main text.

We wish to calculate the laplace beltrami operator of our velocity field $\boldsymbol{v} = \frac{\left(v_0(z)+\delta v^\theta\right)}{r(z)}\left(\frac{\partial}{\partial\theta}\right) + \delta v^z \left(\frac{\partial}{\partial z}\right)$. First we lower the velocity with the metric and act on it with the exterior derivative giving (to linear order)
\begin{equation}
\boldsymbol{d}\boldsymbol{v}^\flat = \left[-r_0\Omega +\partial_zu v_0 - u\Omega+r_0\partial_z\delta v^\theta\right]\mathrm{d}z\wedge\mathrm{d}\theta + \partial_\theta \delta v^z\mathrm{d}\theta\wedge\mathrm{d}z
\end{equation}
next, taking the Hodge star of this and using the asymmetry of the wedge product and the fact that $\text{vol}=\sqrt{|g|}\mathrm{d}\theta\wedge\mathrm{d}z$ we find
\begin{equation}
\star\boldsymbol{d}\boldsymbol{v}^\flat = \Omega -\frac{\partial_zu}{r_0}v_0 -\partial_z\delta v^\theta +\frac{\partial_\theta\delta v^z}{r_0} \text{.}
\end{equation}

Taking the exterior derivative of this leads to
\begin{equation}
\boldsymbol{d}\star\boldsymbol{d}\boldsymbol{v}^\flat = \left[ -\frac{\partial_{zz}u v_0}{r_0} -\partial_{zz}\delta v^\theta +\frac{\partial_{z\theta}\delta v^z}{r_0} -\frac{\partial_zu\Omega}{r_0}\right]\mathrm{d}z + \left[-\frac{\partial_{z\theta}u}{r_0}v_0 -\partial_{z\theta}\delta v^\theta +\frac{\partial_{\theta\theta}\delta v^z}{r_0}\right]\mathrm{d}\theta
\end{equation}
taking the Hodge star of this and applying the inverse metric leads to 
\begin{equation}
\begin{split}
\left(-\star\boldsymbol{d}\star\boldsymbol{d}\boldsymbol{v}^\flat\right)^\sharp = \frac{1}{r_0^2} \left[-\partial_{zz}uv_0 -r_0\partial_{zz}\delta v^\theta +\partial_{z\theta}\delta v^z-\partial_zu\Omega\right]\left(\frac{\partial}{\partial\theta}\right)\\ + \left[ \frac{\partial_{z\theta}uv_0}{r_0^2} +\frac{1}{r_0}\partial_{z\theta}\delta v^\theta -\frac{1}{r_0^2}\partial_{\theta\theta}\delta v^z\right]\left(\frac{\partial}{\partial z}\right)\text{.}
\end{split}
\end{equation}

The contribution to the Stokes equations from the Gaussian curvature is given by
\begin{equation}
-2K\boldsymbol{v} = \frac{2\partial_z^2u}{r_0^2}v_0\left(\frac{\partial}{\partial\theta}\right)
\end{equation}
and from the gradient in the normal velocity we have
\begin{equation}
2(b-2Hg)\cdot\boldsymbol{\nabla} w = 2\left[\left(\begin{matrix}
-\frac{1}{r_0^3} & 0\\
0 & 0
\end{matrix}\right)+\frac{1}{r_0}\left(\begin{matrix}
\frac{1}{r_0^2} & 0\\
0 & 1
\end{matrix}\right)\right]\left(\begin{matrix}
\frac{\partial_\theta w}{r_0}\\
\partial_z w
\end{matrix}\right) = \frac{2}{r_0}\partial_z w\left(\frac{\partial}{\partial z}\right)\text{.}
\end{equation}

Taking Fourier transforms of these we can find the 2D Stokes equations in Fourier space
\begin{equation}
\theta:\quad \eta_m\left[ -\frac{m \tilde{q}}{r^2_0}\bar{\delta v}^z_{q,m} -\frac{\dot{\imath}\tilde{q} \Omega}{r^2_0}\bar{u}_{q,m} + \frac{\tilde{q}^2}{r_0^2} \bar{\delta v}^\theta_{q,m}-\frac{1}{r^3_0}\bar{v_0}\circledast\left[\tilde{q}^2\bar{u}_{q,m}\right]\right]-\frac{
\dot{\imath}m}{r_0}\bar{\delta\sigma}_{q,m}=0
\end{equation}
\begin{equation}
z:\quad \eta_m\left[ \frac{m^2}{r_0^2}\bar{\delta v}^z_{q,m} -\frac{1}{r_0^3}\bar{v_0}\circledast\left[m\tilde{q}\bar{u}_{q,m}\right] - \frac{m\tilde{q}}{r^2_0}\bar{\delta v}^\theta_{q,m} +\frac{2\dot{\imath}\tilde{q}}{r^2_0}\bar{\delta w}_{q,m}\right]-\frac{\dot{\imath}\tilde{q}}{r_0}\bar{\delta\sigma}_{q,m}=0
\end{equation}
where $\circledast$ denotes convolution between the two Fourier transforms in $q$ space. This comes from using the convolution theorem $\mathbb{F}(f\cdot g) = \mathbb{F}(f)\circledast\mathbb{F}(g)$. The continuity equation reads
\begin{equation}
\dot{\imath}m\bar{\delta v}^\theta_{q,m} +\dot{\imath}\tilde{q}\bar{\delta v}^z_{q,m}+\bar{\delta w}_{q,m}=0\text{.}
\end{equation}

From this point it is just a matter of algebra to find the response functions $\bar{\delta v}^\theta_{q,m}$, $\bar{\delta v}^z_{q,m}$ and $\bar{\delta\sigma}_{q,m}$ in terms of $\bar{u}_{q,m}$ and $\bar{\delta w}_{q,m}$.

\begin{equation}
\bar{\delta v}^\theta_{q,m}= \frac{\dot\imath m\left(m^2+3\tilde{q}^2\right)\bar{\delta w}_{q,m} +\dot\imath \tilde{\Omega}\tilde{q}^3\bar{u}_{q,m} +\frac{\tilde{q}^2}{r_0}\bar{v_0}\circledast\left(\tilde{q}^2 \bar{u}_{q,m}\right) - \frac{m\tilde{q}}{r_0}\bar{v_0}\circledast\left(m \tilde{q}\bar{u}_{q,m}\right)}{\left(m^2+\tilde{q}^2\right)^2}
\end{equation}
\begin{equation}
\bar{\delta v}^z_{q,m} = \frac{\dot\imath \tilde{q}\left( \left(\tilde{q}^2 - m^2\right) \bar{\delta w}_{q,m} - m \tilde{q}\Omega \bar{u}_{q,m}\right) +\frac{m^2}{r_0}\bar{v_0}\circledast \left(m\tilde{q}\bar{u}_{q,m}\right) - \frac{m\tilde{q}}{r_0}\bar{v_0}\circledast\left(\tilde{q}^2\bar{u}_{q,m}\right)}{\left(m^2+\tilde{q}^2\right)^2}
\end{equation}
\begin{equation}\label{eq:surfaceTensionLinearResponse}
\bar{\delta\sigma}_{q,m} = \frac{\eta_m\left[\frac{\dot\imath m}{r_0}\bar{v_0}\circledast\left(\tilde{q}^2\bar{u}_{q,m}\right) + \frac{\dot\imath \tilde{q}}{r_0}\bar{v_0}\circledast\left(\tilde{q} m\bar{u}_{q,m}\right) +2\tilde{q}^2\bar{\delta w}_{q,m} - m\tilde{q} \Omega \bar{u}_{q,m}\right]}{\left(m^2+\tilde{q}^2\right)r_0}\text{.}
\end{equation}

We can now make use of the fact that the Fourier transform of the ground-state velocity convolved with some function is given by $\bar{v_0}\circledast\left(\cdot\right) = v_0 - \dot\imath\Omega r_0\partial_{\tilde{q}}\left(\cdot\right)$. Thus we have the following identity
\begin{equation}
\bar{v_0}\circledast\left[f(\tilde{q}\bar{u}_{q,m}\right] = f\left(\tilde{q}\right) \bar{v_0}\circledast\left(\bar{u}_{q,m}\right) -\dot\imath\Omega r_0\bar{u}_{q,m} \partial_{\tilde{q}}f\left(\tilde{q}\right)
\end{equation}

Writing $\bar{\delta w}_{q,m}=\partial_t\bar{u}_{q,m}+\mathcal{O}(u^2)$ we can find a growth rate equation in the shape which is given by the following
\begin{equation}\label{eq:dynamicalEquationDeterministic}
\partial_{\tilde{t}}\bar{u}_{q,m} = -m\frac{\dot\imath v_0\eta_m}{r_0\sigma_0}\bar{u}_{q,m} - \tilde{\Omega} m \partial_{\tilde{q}}\bar{u}_{q,m} + F(q,m)\bar{u}_{q,m}
\end{equation}
where $\tilde{t}=\frac{\sigma_0 t}{\eta_m}$ and
\begin{equation}
F(q,m) = \frac{m\tilde{q}\left(\left(m^2+\tilde{q}^2 \right)^2-2 \tilde{q}^2\right)\tilde{\Omega} - \left(m^2+\tilde{q}^2\right)^2 \left(1+m^4 +\tilde{q}^4 +2 m^2\left(\tilde{q}^2-1\right)\right)}{2\tilde{q}^4}
\end{equation}

\begin{figure}[h!]
\center\includegraphics[width=0.6\textwidth]{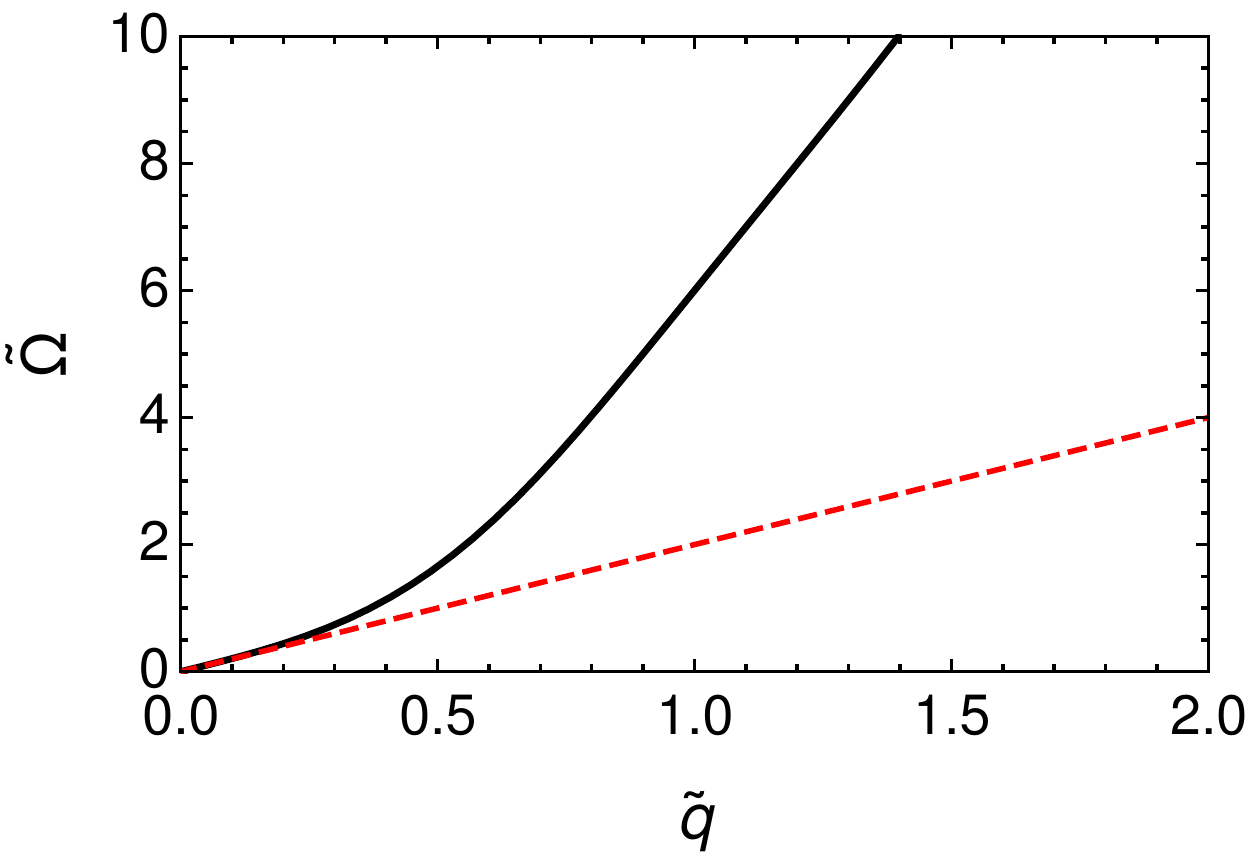}
\caption{\label{fig:stability}Figure showing the zero of $F(q,m=1)$ as a function of $\tilde{q}$ and $\tilde{\Omega}$. The region above the black line is unstable. The dashed red line corresponds to the low $\tilde{q}$ limit of $\tilde{\Omega}\approx 2\tilde{q}$.}
\end{figure}

The solution to this PDE, for with initial data $\bar{u}_{q,m}=u_0\delta(\tilde{q} -\tilde{q}_0)$ is given by 
\begin{equation}\label{eq:pdeSol}
\bar{u}_{q,m}(t) = u_0\delta(\tilde{q}-\tilde{q}_0-m\tilde{\Omega}\tilde{t}) e^{f(q,m)}
\end{equation}
where
\begin{equation}
\begin{split}
f(q,m)=\frac{1}{60} \big[-20 m^2 (2 + 3 \tilde{q}^2) \tilde{t}^3 \tilde{\Omega}^2 +   30 m^3 \tilde{q} \tilde{t}^4 \tilde{\Omega}^3 - 6 m^4 \tilde{t}^5 \tilde{\Omega}^4 -  30 \tilde{t} (5 + 4 \tilde{q}^2 + \tilde{q}^4 - \tilde{q} \tilde{\Omega}) +  15 m \tilde{t} (-4 \frac{\dot\imath v_0 \eta_m}{r_0\sigma_0}\\ +  t (8 \tilde{q} +  4 \tilde{q}^3 - \tilde{\Omega}) \tilde{\Omega}) - \frac{60}{m \tilde{q} \tilde{\Omega} -m^2 \tilde{t} \tilde{\Omega}^2} +  \frac{15}{m} (-\tilde{q}^{-2} + \frac{4}{\tilde{q} \tilde{\Omega}} +  \frac{1}{(\tilde{q} - m \tilde{t} \tilde{\Omega})^2})\big]
\end{split}\text{.}
\end{equation}
and includes the advection of the undulation by the ground-state flow, thus shortening the wavelength of the undulation, i.e.~$\tilde{q}=\tilde{q}(\tilde{t}) = \tilde{q}_0 + m \tilde{\Omega}\tilde{t}$. Because of this advection any $\tilde{q}$ mode which is initially unstable will be advected to stability and the mode will eventually decay. If our system was purely deterministic this would kill the instability in the long time limit, however as the bending rigidity of the membrane has comparable energy to the thermal energy ($\kappa\sim 10k_BT$ where $k_B$ is Boltzmann's constant and $T$ is the temperature) the $\tilde{q}$ spectrum is constantly fed by thermal fluctuations. This means that, to understand the full dynamics of the system we should solve Eq.~\ref{eq:fullGrowthRate} with the addition of a stochastic term describing thermal noise.  

\subsection{Solution to the fluctuation spectrum with thermal noise}
Adding a thermal noise term to Eq.~\ref{eq:fullGrowthRate} we get a Stochastic Partial Differential Equation (SPDE) in $(\tilde{t},\tilde{q})$ given by
\begin{equation}\label{eq:spde}
\partial_{\tilde{t}}\bar{u}_{q,m} = -\tilde{\Omega}m\partial_{\tilde{q}}\bar{u}_{q,m}+ F'(q,m)\bar{u}_{q,m} + \partial_{\tilde{t}}\zeta_{t,q,m}
\end{equation}
where
\begin{equation}
\langle \partial_{\tilde{t}}\zeta_{t,q,m} \partial_{\tilde{t}'}\zeta_{t',q',m'}\rangle = \frac{2k_BT}{\kappa}\frac{\left(m^2+\tilde{q}^2\right)^2}{2\tilde{q}^4}\delta_{q,q'}\delta_{m,m'}\delta\left(\tilde{t}-\tilde{t}'\right) 
\end{equation}
is chosen such that we recover the equipartition result of equilibrium statistical mechanics when $\tilde{\Omega}=\tilde{v}_0=0$. For $\tilde{\Omega}=\tilde{v}_0=0$ this gives the steady state
\begin{equation}
\langle|\bar{u}_{q,1}|^2\rangle = \frac{k_BT}{\kappa\tilde{q}^2\left(\tilde{q}^2+2\right)}\text{.}
\end{equation}

We define $F'(q,m) =F(q,m)-\dot\imath m\tilde{v}_0$. \eq{eq:spde}, with initial data $\bar{u}_{q,m}(0)=u_{0}(q,m)$, is the stochastic version of a Cauchy problem \citep{chow_stochastic_2014}.

In order to solve this SPDE we make use of the method of Stochastic Characteristics \citep{chow_stochastic_2014}. In It\^o form the thermal noise is written as
\begin{equation}
\mathrm{d}\zeta_{t,q,m} = \sqrt{B(\tilde{q},m)}\mathrm{d}W_{t,q,m}
\end{equation}
where $\langle \mathrm{d}W_{t,q,m} \mathrm{d}W_{t',q',m'}\rangle = \delta_{m,m'}\delta_{q,q'}\delta(t-t')\mathrm{d}t\mathrm{d}t'$ and $B(\tilde{q},m)=k_BT\left(m^2+\tilde{q}^2\right)^2/\left(\kappa\tilde{q}^4\right)$.

\eq{eq:spde} is equivalent to the It\^o integral
\begin{equation}
\bar{u}_{q,m}(\tilde{t}) = u_0(\tilde{q},m) -m \tilde{\Omega} \int_0^{\tilde{t}}\partial_{\tilde{q}}\bar{u}_{q,m}(s)\mathrm{d}s + \int_0^{\tilde{t}}\left[F'(q,m)\bar{u}_{q,m}(s)\mathrm{d}s + \mathrm{d}\zeta_{s,q,m}\right]\text{.}
\end{equation}

In order to solve this we introduce the following characteristics
\begin{equation}
\begin{split}
&\phi_{t}(q) = \tilde{q} +m \tilde{\Omega} \int_0^{\tilde{t}}\mathrm{d}s = \tilde{q} +m \tilde{\Omega}\tilde{t}\\
&\eta_{t}(q,r) = r + \int_0^{\tilde{t}}\eta_s \left(q,r\right)F'\left(\phi_s(q),m\right)\mathrm{d}s + \int_0^{\tilde{t}} \sqrt{B(\phi_s(q),m)}\mathrm{d}W_{t,q,m}
\end{split}
\end{equation}
where $r$ is some stochastic initial condition to the stochastic characteristic curve $\eta(q,r)$. $\phi_t(q)$ is just the standard deterministic characteristic associated with a linear transport equation describing a translation of the wavenumber, $\tilde{q}$, in time. The solution to these stochastic integral equations has a unique solution (for certain regularity conditions on the noise), this solution defines a stochastic flow of diffeomorphism. This leads to the solution to the stochastic Cauchy problem, posed by \eq{eq:spde} and it's initial data, for full details see Ref.~\cite{chow_stochastic_2014}. In our case the solution is given by
\begin{equation}\label{eq:uStoch}
\begin{split}
\bar{u}_{q,m} = &u_{0}\left(\phi_t^{-1}(q),m\right)\exp\left[\int_0^{\tilde{t}}F'(\phi_s(y),m)\mathrm{d}s\right]\Bigg\vert_{y=\phi^{-1}_t(q)}\\
&+ \int_0^{\tilde{t}} \exp\left[\int_\tau^{\tilde{t}}F'(\phi_s(y),m)\mathrm{d}s\right]\sqrt{B( \phi_{\tau}(y),m)}\mathrm{d}W_{\tau}|_{y=\phi^{-1}_t(q)}
\end{split}
\end{equation}
which does not depend explicitly on the characteristic $\eta_t(q,r)$ due to the relatively simple form of our SPDE.

Evaluating the integrals in the exponentials we find
\begin{equation}
\begin{split}
&f'(q,m,t)=\int F'(\phi_t(y),m)\mathrm{d}t\\ &= \Bigg[-15 m^5 \Omega  (m t \Omega +q)-40 m^2 (m t \Omega +q)^6\\
&+60 m \left(m^2-1\right) \Omega  (m t \Omega +q)^3 \log (m t \Omega +q)-30 \left(6 m^4-2 m^2+1\right) (m t \Omega +q)^4\\
&+10 m^4 \left(m^2-1\right)^2+60 \left(2 m^6-2 m^4+m^2\right) (m t \Omega +q)^2-6 (m t \Omega +q)^8\\
&+15 m \Omega  (m t \Omega +q)^5\Bigg]\bigg(60 m \Omega  (m t \Omega +q)^3\bigg)^{-1} -\dot\imath m\tilde{v}_0 t\\ &= f(q,m,t) -\dot\imath m\tilde{v}_0 t\text{.}
\end{split}
\end{equation}

We want to consider the steady state of the fluctuations at a time when any dependence on this initial data has decayed so by taking the complex conjugate of \eq{eq:uStoch} squared and averaging we find
\begin{equation}
\langle|\bar{u}_{q,m}|^2(\tilde{t})\rangle = e^{2f(\tilde{q}-m\tilde{\Omega}\tilde{t},m,\tilde{t})}\int_0^{\tilde{t}}B\left(\tilde{q}+m\tilde{\Omega} \left(\tau-\tilde{t}\right),m\right)e^{-2f(\tilde{q}-m\tilde{\Omega}\tilde{t},m,\tau)}\mathrm{d}\tau
\end{equation}
where we have input the characteristic curves and their inputs explicitly and neglected the term describing the dynamics of the initial data as we are only interested in the steady state.

If we consider the case of the $m=1$ mode then the equilibrium fluctuations are known to be critical in the $\tilde{q}\to 0$ limit \citep{fournier_critical_2007}. Because of this we introduce a cut of wavenumber $\tilde{q}_0$ that corresponds to the length-scale of the longest fluctuation on the finite tube. This implies that the noise kernel of our system has only localized support on the interval $\tau-\tilde{t}\in[\frac{\tilde{q}_0-\tilde{q}}{\Omega},0]$, so we can use this to truncate the limits of our integration. Thus, the $m=1$ steady state fluctuations are given by
\begin{equation}\label{eq:steadyFluct}
\langle|\bar{u}_{q,1}|^2\rangle = e^{2f(\tilde{q}-\tilde{\Omega}\tilde{t},1,\tilde{t})}\int_{\frac{\tilde{q}_0-\tilde{q}}{\Omega}+\tilde{t}}^{\tilde{t}}B\left(\tilde{q}+\tilde{\Omega} \left(\tau-\tilde{t}\right),1\right)e^{-2f(\tilde{q}-\tilde{\Omega}\tilde{t},1,\tau)}\mathrm{d}\tau
\end{equation}
which, after performing the integration, does not depend on $\tilde{t}$.

\begin{figure}
\center\includegraphics[width=0.8\textwidth]{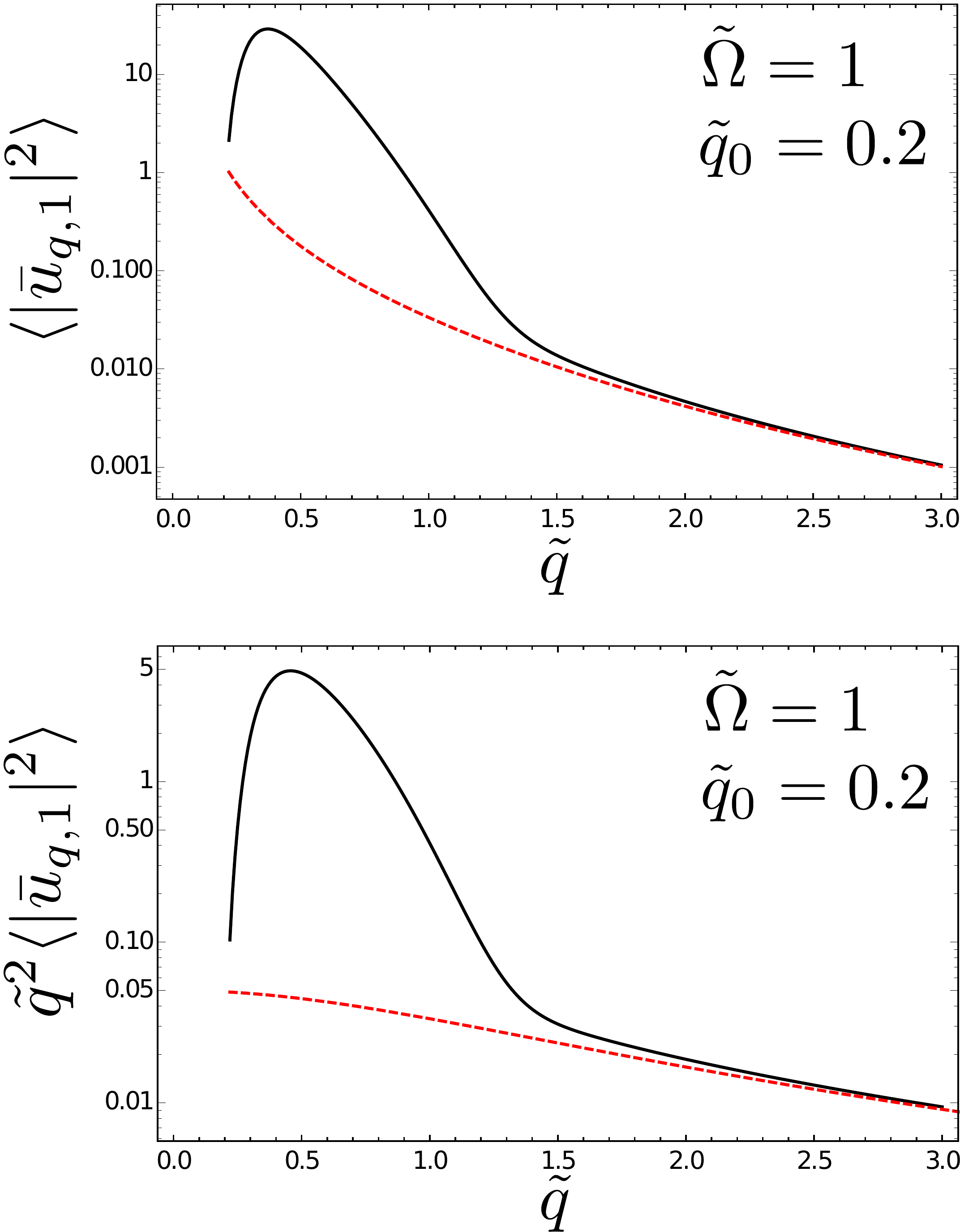}
\caption[Stead state fluctuations on sheared tube]{\label{fig:fluctuations}The steady state fluctuations for a sheared tube (\eq{eq:steadyFluct}) and their Fourier ``gradients'' ($\tilde{q}^2\langle|\bar{u}_{q,1}|^2\rangle$) with shear rate, $\tilde{\Omega}=1$, and long wavelength cut-off, $\tilde{q}_0=0.2$. We choose the bending rigidity to be $\kappa=10 k_BT$. The dashed red line shows the equivalent thermal fluctuations.}
\end{figure}

The steady states of $\langle|\bar{u}_{q,1}|^2\rangle$ and the $z$ part of their spacial gradients in Fourier space $\tilde{q}^2\langle|\bar{u}_{q,1}|^2\rangle$ are plotted in \fig{fig:fluctuations} for $\tilde{\Omega}=1$ and $\tilde{q}_0=0.2$. We also plot the equivalent thermal fluctuations given by
\begin{equation}
\langle|\bar{u}_{q,1}|^2\rangle = \frac{k_BT}{\kappa\tilde{q}^2\left(\tilde{q}^2+2\right)}\text{.}
\end{equation}
which we plot as a red dashed line.

We want to find a criterion for when the linearisation breaks down. We choose a proxy for this to be when the spacial gradients of the steady state are large,
\begin{equation}\label{eq:linearisationCondition}
C\langle|\nabla u(z,\theta)|^2\rangle=1\text{,}
\end{equation}
where we set $C=1,4$ to correspond to $32\%$ and $5\%$ of fluctuations breaking the linearisation respectively. To calculate this we write the following
\begin{equation}
\nabla u = \sum_{m}\int\frac{\mathrm{d}q}{2\pi} \bar{u}_{q,m}\left(\begin{matrix}
\dot\imath q\\
\dot\imath \frac{m}{r_0}
\end{matrix}\right)e^{\dot\imath q z+\dot\imath m \theta}\text{.}
\end{equation}
By taking the Hermitian conjugate of this we find $\left(\nabla u \left(z',\theta'\right)\right)^\dag$ which can be written as a sum over $q'$ and $m'$. By integrating over $q'$ and summing over $m'$, then setting $z=z'$ and $\theta=\theta'$ we find
\begin{equation}
\langle|\nabla u|^2\rangle = \sum_{m}\int\frac{\mathrm{d}\tilde{q}}{2\pi}\left(\tilde{q}^2+m^2\right)\langle|\bar{u}_{q,m}|^2\rangle
\end{equation}
where we have moved back to dimensionless units used in the main paper. As the largest contribution to the fluctuations comes from the $m=1$ mode we write
\begin{equation}
\langle|\nabla u|^2\rangle \approx \int_{\tilde{q}_0}^{\infty}\frac{\mathrm{d}\tilde{q}}{2\pi}\left(\tilde{q}^2+1\right)\langle|\bar{u}_{q,1}|^2\rangle
\end{equation}
which is the expression we compute numerically (taking the upper limit to be $\tilde{q}=10$) to find the linearisation condition, \eq{eq:linearisationCondition}.

\subsection*{Scaling analysis for the critical frequency in small $\tilde{q}$ regime}
For $q\sim 1/L$, the first order correction to the curvature scales like $H\sim \frac{u}{L^2}$ so that the elastic force-per-unit-area scales like $f_{\text{el}}\sim\frac{u\sigma_0}{L^2}$, while the off-diagonal components of the second fundamental form scale like $b\sim \frac{u}{r_0 L}$ and hence the viscous force-per-unit-area scale like $f_{\text{vis}}\sim\frac{\eta_m\Omega u}{r_0 L}\sim\frac{u \eta_m  \nu}{L^2}$. Balancing these forces gives a critical frequency 
\begin{equation}
\nu_{\text{crit}}\sim\frac{\sigma_0}{\eta_m}\text{.}
\end{equation}

\subsection*{Surface tension fluctuations and possible scission by membrane lysis}
One possible mechanism for membrane tube scission involves lysis of the membrane due to increases in surface tension. The  surface tension fluctuations at the linear level may be relevant here, although the largest growth in surface tension may be in the nonlinear regime.

Making use of \eq{eq:surfaceTensionLinearResponse} and substituting for the normal velocity using \eq{eq:dynamicalEquationDeterministic} we can write the variation in surface tension (in Fourier space) as
\begin{equation}
\frac{\delta\bar{\sigma}_{m,q}}{\sigma_0} = \frac{2\left(\tilde{\Omega}\tilde{q}m+\tilde{q}^2 F(q,m)\right)}{m^2+\tilde{q}^2}\bar{u}_{q,m}
\end{equation}
and taking the square average we find an estimate for the surface tension fluctuations
\begin{equation}
\frac{\langle|\delta\bar{\sigma}_{m,q}|^2\rangle}{\sigma_0^2} = \frac{4\left(\tilde{\Omega}\tilde{q}m+\tilde{q}^2 F(q,m)\right)^2}{\left(m^2+\tilde{q}^2\right)^2}\langle|\bar{u}_{q,m}|^2\rangle\text{.}
\end{equation}

\begin{figure}
\includegraphics[width=0.5\textwidth]{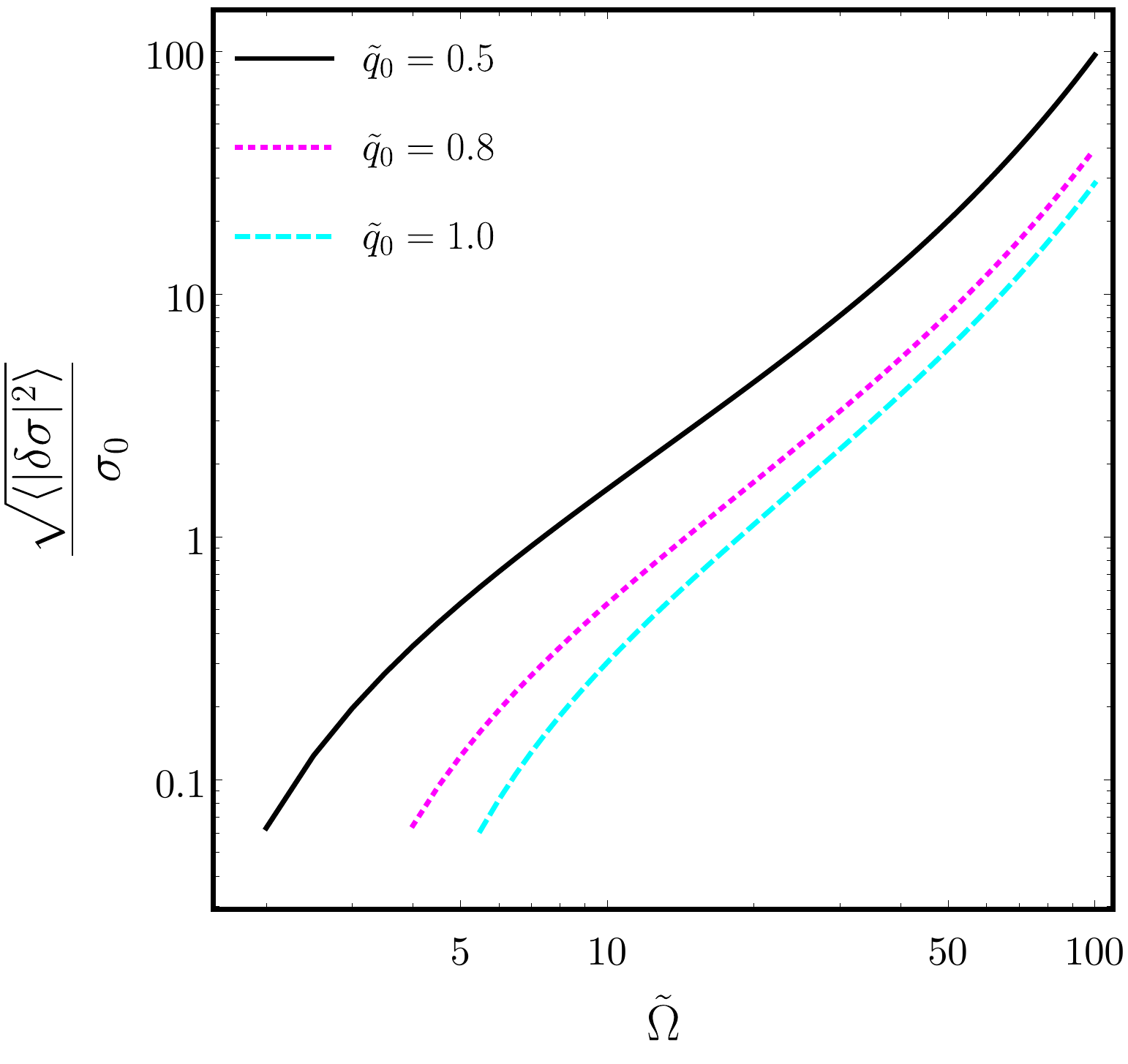}
\caption{\label{fig:surfaceTensionFluctuations}The RMS of the surface tension fluctuations, $\frac{\sqrt{\langle|\delta\sigma|^2\rangle}}{\sigma_0}$ against dimensionless shear, $\tilde{\Omega}$, plotted for  various cut-off wavelengths, $\tilde{q}_0$.}
\end{figure}

If we then want to know the real-space surface tension fluctuations we can invert the Fourier transform
\begin{equation}
\frac{\langle|\delta\sigma|^2\rangle}{\sigma_0^2}=\int_{\tilde{q_0}}^{\Lambda}\frac{\langle|\delta\bar{\sigma}_{m,q}|^2\rangle}{\sigma_0^2}\frac{\mathrm{d}\tilde{q}}{2\pi}
\end{equation}
here $\Lambda$ is a UV cut-off introduced to stop the divergence of $\langle|\delta\bar{\sigma}_{m,q}|^2\rangle$ at high $\tilde{q}$. This divergence is due to the dependence of the surface tension fluctuations on the normal velocity fluctuations, which diverge at high $\tilde{q}$. The formal way to treat this would be with renormalization group methods however, as at high $\tilde{q}$ the fluctuations behave like the equilibrium fluctuations, for simplicity we chose $\Lambda$ to be given by $\tilde{\Omega}\Lambda m+\Lambda^2 F(\Lambda,m)=0$ which gives the contribution that is dominated by the shear-driven (non-equilibrium) fluctuations. The rms value of the surface tension fluctuations are plotted as a function of $\tilde{\Omega}$ for values of $\tilde{q}_0\sim 1$ in \fig{fig:surfaceTensionFluctuations}. Note that for the values associated with high effective viscosity similar to that measured in live cell membrane tubes \cite{brochard-wyart_hydrodynamic_2006} (high $\tilde{\Omega}$) can lead to an order of magnitude increase (or more) in surface tension fluctuations, maximal on the outside of the helix. These fluctuations could be sufficient to play a role in membrane lysis.

\subsection*{Notes on screening by bulk flows}
We will consider hydrodynamics on a static membrane tube (i.e. we assume that the cylindrical geometry is stable to perturbations in shape). In the limit of small inertia the $3$D velocity field, $\vec{u}$, satisfies the continuity and Stokes equations
\begin{equation}\label{eq:3dStokes}
\vec{\nabla}\cdot\vec{u}=0;\quad \eta\nabla^2\vec{u}=\vec{\nabla}P
\end{equation}
where $P$ is the pressure and $\eta$ the viscosity. This is coupled to the membrane velocity at the boundary with a no-slip condition.

Stress balance at the membrane is imposed by the $2$D continuity and Stokes equations and, for surfaces of zero Gaussian curvature, can be written as
\begin{equation}\label{eq:2dStokesSimple}
\nabla_i v^i=0; \quad \eta_m\Delta_{\text{LB}}v_i -\nabla_i\sigma=t_i^{+}+ t_i^-
\end{equation} 
where $\eta_m$ is the ($2$D) membrane viscosity, $\sigma$ is the surface tension, $\boldsymbol{v}=v^i\boldsymbol{e}_i$ is the tangential membrane velocity and $\Delta_{\text{LB}}$ is the Laplace-Beltrami operator (formally this corresponds to $\Delta_{\text{LB}}=\boldsymbol{\delta}\boldsymbol{d}$ where $\boldsymbol{d}$ is the exterior derivative and $\boldsymbol{\delta}$ is the co-differential). The combined operator $\boldsymbol{\delta}\boldsymbol{d}$ is the generalization of the curl-curl operator to a manifold and acts like a Laplacian \cite{rahimi_curved_2013,arroyo_relaxation_2009}. The symbols $t_i^\pm$ are the traction forces from the bulk fluid acting on the membrane ($\pm$ denoting interior and exterior respectively)\cite{arroyo_relaxation_2009,fournier_hydrodynamics_2015}.

We will consider a system of a membrane tube with radius $r_0=\sqrt{\frac{\kappa}{2\sigma_0}}$, where $\kappa$ is the bending rigidity of the membrane and $\sigma_0$ is the equilibrium surface tension. This is the radius which minimizes the Helfrich Hamiltonian for a fluid membrane
\begin{equation}
\mathcal{F} = \int_{\Gamma}\mathrm{d}A_{\Gamma} \left(2\kappa H^2 +\sigma_0\right)
\end{equation} 
where $\Gamma$ and $\mathrm{d}A_{\Gamma}$ denote the manifold describing the neutral surface of the membrane and its associated area element, and $H$ is the mean curvature \cite{zhong-can_bending_1989}. For typical membrane tubes fissioned by dynamin $r_0\approx 10\text{nm}$ \cite{roux_reaching_2014}.

We use standard cylindrical coordinates $\left(r,\theta,z\right)$ and take the boundary condition for flow on the membrane to be $\boldsymbol{v}|_{z=0}=v_0\vec{e}_\theta$, we treat this as an approximation to the flow induced by dynamin.

\begin{figure}
\includegraphics[width=0.45\textwidth,trim= 3cm 9cm 3.5cm 9.5cm, clip=true]{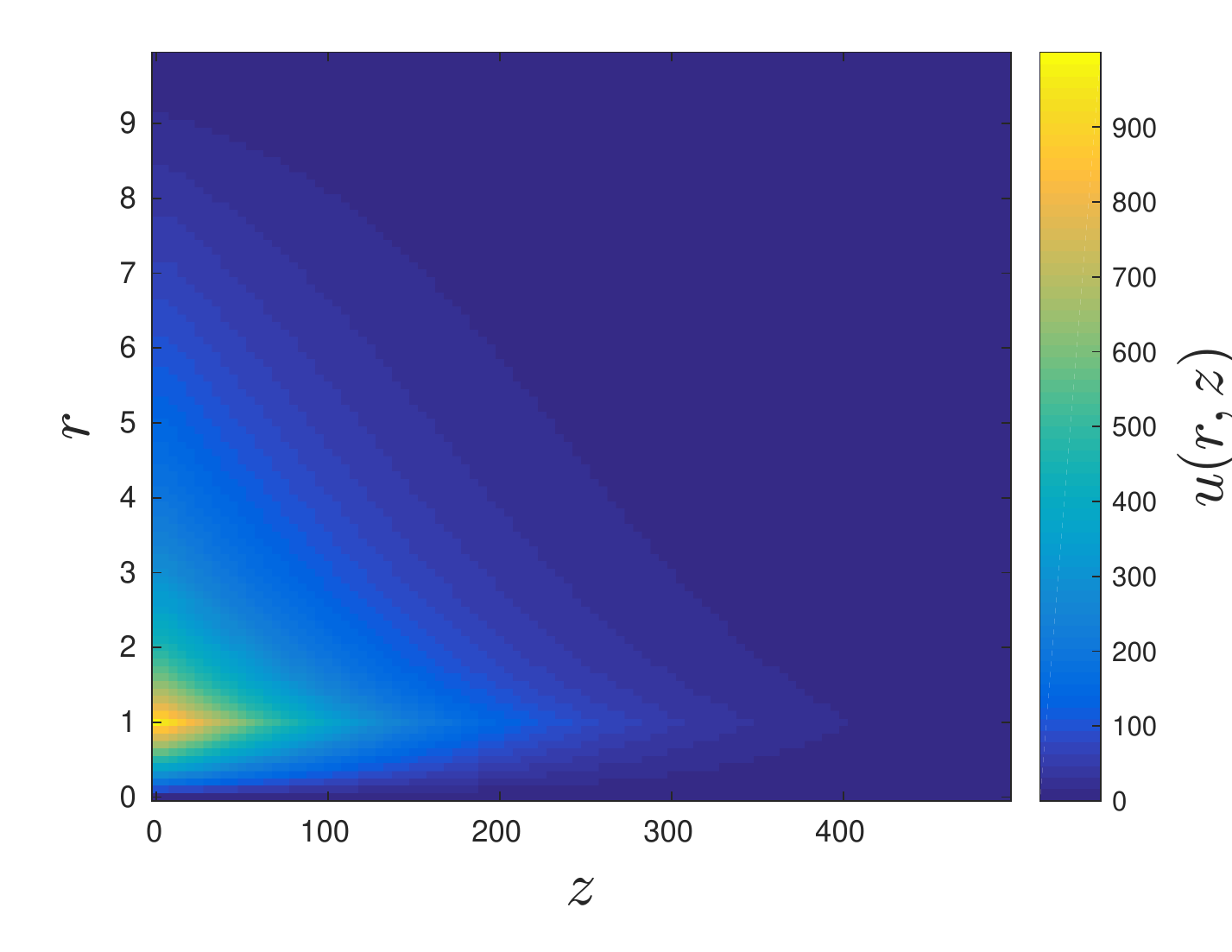}
\caption{\label{fig:flowField}Flow field for the ground-state of the spinning membrane tube with radius $r_0=1.0$, and Saffman-Delbr\"uck length $\frac{L_{\text{SD}}}{r_0}=\frac{\eta_{m}}{\eta r_0}=10^4$. The boundary condition on the tube at $z=0$ is $v(0)=v_0$ where $\frac{v_0}{r_0}=10^3\text{s}^{-1}$.}
\end{figure}

We can then solve the \eqref{eq:3dStokes} \& \eqref{eq:2dStokesSimple}, making use of symmetry $\boldsymbol{v}=v(z)\vec{e}_\theta$, $\vec{u}=u(r,z)\vec{e}_\theta$ they reduce to
\begin{equation}
\begin{split}
\frac{1}{r}\partial_r\left(r\partial_ru_\theta\right) +\partial_z^2u_\theta -\frac{u_\theta}{r^2}=0\\
\eta_M\partial_z^2v + t_\theta^+ + t_\theta^-=0
\end{split}
\end{equation}
where $t_\theta^\pm = \lim_{r\to r_0}\eta r \partial_r\left(\frac{\partial_r u^\pm}{r}\right)$. We can now solve this numerically by direct methods (taking a Neumann boundary condition for the bulk flow at $z=0$ and $u=0$ at large distance and $r=0$) \cite{ferziger_computational_2002}. The flow field computed by this method can be seen in \fig{fig:flowField}.

\begin{figure}
\center\includegraphics[width=0.45\textwidth,trim= 3cm 9cm 3.5cm 9.5cm, clip=true]{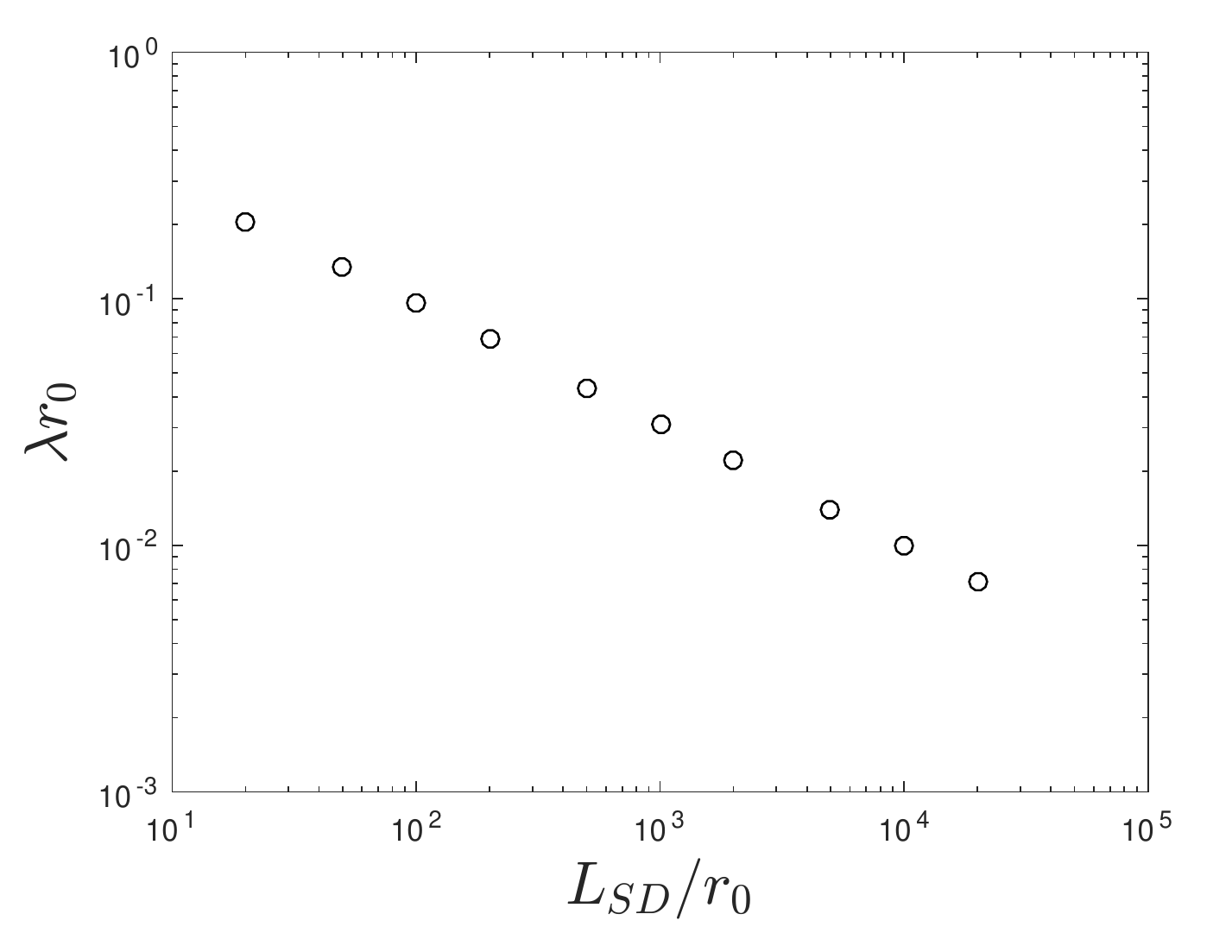}
\caption{\label{fig:screeningLength}Flow-field decay rate, $\lambda$ (with units $\text{Length}^{-1}$) against Saffman-Delbr\"uck length $L_{\text{SD}}$ for tube spinning velocity at $z=0$ given by $\frac{v_0}{r_0}=10^3\text{s}^{-1}$.}
\end{figure}

To understand how the flow field on the membrane varies with Saffman-Delbr\"uck length it is helpful to examine the analytic solutions to the coupled membrane bulk system in Fourier space. The flow field on the membrane in response to a point force in the $\theta$ direction, $F_\theta$, was found analytically by Henle \& Levine \cite{henle_hydrodynamics_2010}, and in the limit $r_0\ll L_{\text{SD}}$ this gives
\begin{equation}
\boldsymbol{v}\approx v_0\vec{e}_\theta\exp\left[-\frac{\sqrt{2}|z|}{\sqrt{L_{\text{SD}}r_0}}\right]\text{.}
\end{equation}
In the original paper our boundary condition corresponds to $v_0 = \frac{F_\theta}{4\pi\eta_m}\sqrt{\frac{L_{\text{SD}}}{2 r_0}}$. Note that this is $\theta$ independent as the $m=0$ Fourier mode dominates the bulk dynamics in this limit, so each cross-section of the tube rotates with a constant velocity. This means that the flow on a tube is screened like $v\sim e^{-\lambda|z|}$ where $\lambda=\frac{\sqrt{2}}{\sqrt{L_{\text{SD}}r_0}}$. This approximate analytical expression can be compared to numerical solutions where we find that it reproduces the correct power law relation between $\lambda$ and $L_{\text{SD}}$, see \fig{fig:screeningLength}.

For flows with large $L_{\text{SD}}/r_0\sim 10^{3}-10^4$ this gives a screening length of order $100r_0$ so as long as we consider flows where $L\lesssim 10r_0$ then membrane dissipation should dominate.

\subsection*{Propagation of ground-state shear flow on the tube}
In order to justify using the steady state low Reynolds hydrodynamics equations on the lipid membrane we briefly discuss the time-scales on which we would expect this approach to break down. One would expect the flow to propagate along the tube at momentum diffusion time-scale $L^2h\rho/\eta_m$ where $\rho\sim 10^{3}\text{kg m}^{-3}$ is the density of lipids and $h\sim 10^{-9}\text{m}$ the height of the membrane. This gives a time of propagation of the shear flow along a tube of $10\mu\text{m}$ to be $\tau_{\text{bulk}}\sim 10^{-7}\text{s}$. These time-scales are much faster than the dynamics we are analysing so it is sufficient to consider the steady state Stokes equations for our purposes. Another possible inertial time-scale that might be of relevance for the instability dynamics is that of the sound mode $c\sim \sqrt{\frac{K}{\rho h}}$ where $K\sim0.1 \text{N m}^{-1}$ is the $2$D Bulk modulus. However this gives a time-scale at a similar order of magnitude, so we assume this can be neglected.

\subsection*{Effects of more realistic geometry}
To try and understand the effect of the instability in more complex geometry (in particular with non-zero Gaussian curvature in the ground state), we need to consider the term driving the instability as the full calculation becomes intractable very quickly. All the forces acting normal to the membrane which drive the instability are due to the term $b^{i}{}_j\nabla_i v^j$, in particular the driving force (per area) is set by the linear response coefficient of the mixed second derivative of the shape, $k_{\theta z}(z)$ which is now a function of $z$ due to change in geometry (specifically the non-constant gradient in the flow field ground state). The driving force per unit area scales like
\begin{equation}
f_{\text{driving}}\sim 2\eta_m k_{\theta z}(z)\frac{\partial^2 u}{\partial\theta\partial z}
\end{equation}
so we will consider how $k_{\theta z}(z)$ changes as we change the geometry of our ground-state.

For some general axisymmertic ground-state parametrized by the vector $\vec{X}=\left(r(z)\cos\theta,r(z)\sin\theta,z\right)$ with ground-state flow field $v0(z)\vec{e}_\theta$ we find (up to linear order in perturbations)
\begin{equation}
b^{i}{}_j\nabla_i v^j= a_{z0\theta 0} \delta v_{z} + a_{z1\theta 0} \partial_z\delta v_{z}+ k_{\theta z}\frac{\partial^2 u}{\partial\theta\partial z} + b_{z0\theta 1} \partial_\theta\delta v_{\theta} +k_\theta \partial_\theta u
\end{equation}
where 
\begin{equation}
\begin{split}
a_{z0\theta 0} &= \frac{-r'(z) - 2 r'(z)^3 - r'(z)^5 +  r(z)^2 r'(z) r''(z)^2}{
r(z)^2 (1 + r'(z)^2)^{5/2}}\\
a_{z1\theta 0} &= \frac{r''(z)}{(1 + r'(z)^2)^{3/2}}\\
 k_{\theta z} &= \Big[-v0(z) r'(z) - v0(z) r'(z)^3 + r(z) v0'(z) + 
 r(z) r'(z)^2 v'(z) + r(z) v0(z) r'(z) r''(z)\Big]\\
 &\times \left(r(z)^2(1 + r'(z)^2)^{5/2}\right)^{-1}\\
 b_{z0\theta 1} &= \frac{1}{r(z)^2\sqrt{1+r'(z)^2}}\\
 k_\theta &= \frac{v0(z)}{r(z)^3\sqrt{1+r'(z)^2}}
\end{split}
\end{equation}

\subsubsection*{Neck (Catenoid)}
\begin{figure}
\includegraphics[width=0.2\textwidth,trim = 0cm 3cm 0cm 3cm,clip=true]{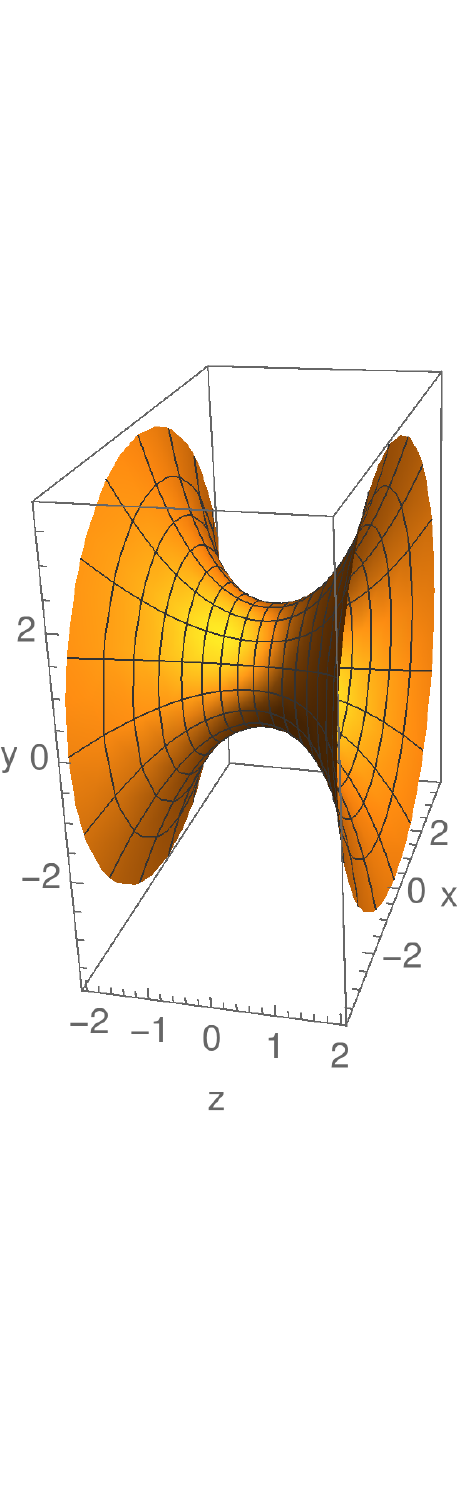}
\includegraphics[width=0.6\textwidth]{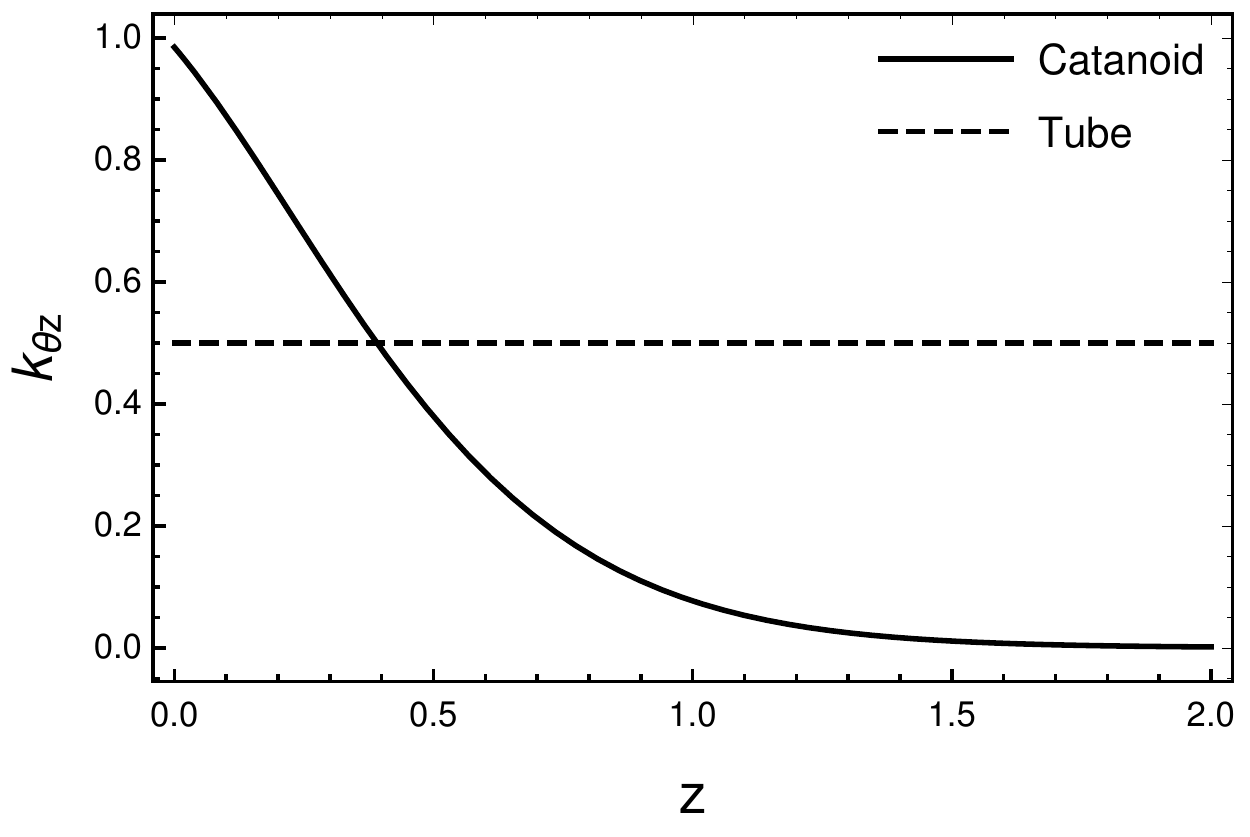}
\caption{\label{fig:catanoid}Plot of the catenoid and the linear response coefficient for the helical shape perturbations on such a surface.}
\end{figure}

To consider the effect of the instability in a more realistic \textit{in-vivo} situation, for example on the neck of a budding vesicle, we look at the ground state flows and $k_{\theta z}$ on a catenoid, $r(z)= r_0\cosh\left(\frac{z}{r_0}\right)$. The ground state surface flow is solved numerically with boundary conditions $v(0)=1$, $v(2)=0$ taking $r_0=1$ and $L=2$ for simplicity. From this we can evaluate $k_{\theta z}$ and compare to the case of a tube.  This is shown in \fig{fig:catanoid}. Note the amplification of $k_{\theta z}$ by a factor of $2$ near the centre of the catenoid when compared to the tube.  The consequences of this for dynamin are discussed in the main paper.

\end{document}